\documentclass{article}
\usepackage[utf8]{inputenc}
\usepackage[left=2.5cm,top=3cm,right=2.5cm,bottom=3cm]{geometry}
\usepackage[colorlinks=true, pdftex, citecolor=red, urlcolor=blue]{hyperref}
\usepackage{amsmath,xspace,graphicx,floatrow,cite}
\pdfoutput=1
\title{Helicity Antenna Showers for Hadron Colliders}
\author{Nadine Fischer$\,^a$, Andrew Lifson$\,^{a,b}$, Peter Skands$\,^a$}
\def\newinstitute{$^a$ {\small School of Physics and Astronomy, Monash University, Clayton, 
VIC 3800, Australia} \\
$^b$ {\small ETH Z{\"u}rich, Z{\"u}rich, CH 8092, Switzerland}}
\date{\today}
\makeatletter
\let\newtitle\@title
\let\newauthor\@author
\let\newdate\@date
\makeatother

\newcommand{\mrm}[1]{\mathrm{#1}}
\newcommand{\mc}[1]{\mathcal{#1}}

\newcommand{\ant}[1]{\mc{A}\left(#1\right)}
\newcommand{\antIdx}[1]{\mc{A}_{#1}}

\newcommand{\Phis}[1]{{\scriptstyle{\Phi_{#1}}}}
\newcommand{\Phisp}[1]{{\scriptstyle{\Phi_{#1}'}}}

\newcommand{\tsc}[1]{\textsc{#1}}
\newcommand{\ttt}[1]{\texttt{#1}}

\newcommand{\vc}{\tsc{Vincia}\xspace}
\newcommand{\py}{\tsc{Pythia}\xspace}
\newcommand{\mg}{\tsc{Madgraph}\xspace}

\newcommand{\riv}{\tsc{Rivet}\xspace}
\newcommand{\fj}{\tsc{Fastjet}\xspace}
\newcommand{\hw}{\tsc{Herwig}\xspace}

\newcommand{\pw}{\tsc{Powheg}\xspace}
\newcommand{\js}{\tsc{Jetset}\xspace}

\newcommand{\eqRef}[1]{eq.~\eqref{#1}\xspace}

\newcommand{\eqsRef}[1]{eqs.~\eqref{#1}\xspace}

\newcommand{\appRef}[1]{app.~\ref{#1}\xspace}
\newcommand{\AppRef}[1]{App.~\ref{#1}\xspace}
\newcommand{\appsRef}[1]{apps.~\ref{#1}\xspace}

\newcommand{\figRef}[1]{fig.~\ref{#1}\xspace}
\newcommand{\FigRef}[1]{Fig.~\ref{#1}\xspace}

\newcommand{\secRef}[1]{sec.~\ref{#1}\xspace}

\newcommand{\tabRef}[1]{tab.~\ref{#1}\xspace}

\newlength{\rivetLength}
\newcommand{\rivetFigure}[1]{
  \includegraphics[width=\rivetLength]{#1}
}
\newcommand{\rivetFigureRatio}[1]{
  \includegraphics[width=\rivetLength]{#1}\vspace*{-0.5mm}
}

\newfloatcommand{capbtabbox}{table}[][\FBwidth]

\newenvironment{newItem}
{\begin{list}{$\bullet$}{
  \setlength{\topsep}{2mm}\setlength{\partopsep}{1mm}
  \setlength{\itemsep}{0.5mm}\setlength{\parsep}{1mm}
}}
{\end{list}}

\begin{document}

\begin{flushright}
CoEPP-MN-17-5 \\
MCNET-17-13
\end{flushright}
\vspace{5mm}
\begin{flushleft}
{\LARGE\bf \newtitle}
\\[10mm]
{\Large \newauthor}
\\[5mm]
{\newinstitute}
\end{flushleft}
\vspace{5mm}

\begin{abstract}
We present a complete set of helicity-dependent $2\to 3$ antenna functions for  QCD initial- and final-state radiation. 
The functions are implemented in the \vc shower Monte Carlo framework and are used to generate showers for hadron-collider processes in which helicities are explicitly sampled (and conserved) at each step of the evolution. Although not capturing the full effects of spin correlations, the explicit helicity sampling  does permit a significantly faster evaluation of fixed-order matrix-element corrections.
A further speed increase is achieved via the implementation of a new fast library of analytical MHV amplitudes, while matrix elements from \mg are used for non-MHV configurations. A few examples of applications to QCD $2\to 2$ processes are given, comparing the newly released \vc 2.200 to \py 8.226.
\end{abstract}

\graphicspath{{images/}}

\section{Introduction}
\label{sec:intro}

The description of bremsstrahlung processes in parton-shower event generators typically starts from the probability density for unpolarised partons to emit unpolarised radiation, i.e., DGLAP kernels or dipole/antenna functions summed over outgoing and averaged over incoming polarisations/helicities. 
One way of incorporating nontrivial polarisation effects, used in \py~\cite{Sjostrand:2014zea}, is to correlate the plane in which a gluon is produced, with the plane in which it subsequently branches, taking linear-polarisation effects into account on the intermediate propagator, and casting the result in terms of a non-uniform selection of the azimuthal $\varphi$ angle around the direction of the branching gluon, see, e.g.,~\cite{Webber:1986mc}. A more complete, but also more cumbersome, alternative, used in \hw~\cite{Bellm:2017bvx}, is to keep track of spin correlations explicitly, using a spin-density matrix formalism~\cite{Knowles:1988vs,Knowles:1988hu,Richardson:2001df}. In both cases, the nontrivial angular correlations ultimately arise from dot products between  reference vectors expressing \emph{linear} polarisations. 

By contrast, a \emph{helicity} basis 
does not rely on any external reference vectors, and hence helicity-dependence in and of itself does not  generate any nontrivial angular correlations. Nonetheless, helicity-dependent radiation functions, as used for final-state radiation in \vc for a few years~\cite{Larkoski:2013yi}, do have some advantages: helicity conservation can be made explicit, allowing to trace helicities through the shower; unphysical helicity configurations are prevented from contributing to sums and averages; and the explicit helicity assignments allow faster evaluations of matrix-element correction (MEC) factors, since only a single (or a few) helicity amplitudes need to be evaluated for each ME-corrected parton state~\cite{Larkoski:2013yi}.

The  concept of ME corrections was first developed to improve the description of radiation in \py (then called \js) outside the collinear region to agree with first-order matrix elements for $e^+e^-\to3\,\mathrm{jets}$~\cite{Bengtsson:1986hr,
Bengtsson:1986et}, and was since extended to correct the first emission in a wide range of resonance-decay processes and some (colour-singlet) production processes~\cite{Miu:1998ju,Norrbin:2000uu}. It was also used as a component of the first ME correction strategies in \hw~\cite{Seymour:1994df,Corcella:1998rs}, and it 
forms the basis of the treatment of real corrections within the \pw formalism~\cite{Nason:2004rx,Frixione:2007vw}. We note that, in these approaches, only the first shower emission is corrected, essentially by applying a multiplicative factor,
\begin{equation}
{\cal R}^\mathrm{MEC} = \frac{\mathrm{ME}}{\mathrm{PS}}~,
\end{equation}
 to the shower kernels, where $\mathrm{ME}$ is the relevant matrix-element expression (typically called ``$R$'' in \pw notation) and $\mathrm{PS}$ represents the (sum of) parton-shower contributions to the given phase-space point. 

The limitation to single emissions was lifted by the development of iterated ME corrections~\cite{Giele:2011cb}\footnote{We note that a form of iterated ME corrections is also used throughout the \py\ showers to impose quark-mass corrections~\cite{Norrbin:2000uu}, but the resulting process-dependent nonsingular terms will still only be fully correct for the first emission.}, implemented in \vc~\cite{Giele:2007di,Fischer:2016vfv}, again first in the context of $e^+e^-\to\mathrm{jets}$~\cite{Giele:2011cb} and subsequently for hadron collisions~\cite{Fischer:2016vfv,Fischer:2017yja}. Importantly, the most recent study in~\cite{Fischer:2017yja} extended the formalism to strongly-ordered and non-Markovian shower algorithms, expanding its applicability to essentially any shower algorithm in modern MC generators. 
Although a helicity-dependent (and hence computationally faster) version of the iterated-MEC algorithm was developed for final-state radiation~\cite{Larkoski:2013yi}, a fully-fledged helicity-dependent version for hadron collisions (and for strongly-ordered non-Markovian showers) has so far been missing. The aim of this paper is to develop this missing piece, while simultaneously presenting a complete set of  helicity-dependent (and positive-definite) antenna functions for $2\to 3$ branchings for both initial- and final-state radiation.
In addition, some helicity configurations (called ``maximally helicity violating'') can be expressed in  compact analytical forms, hence we use such amplitudes for QCD $2\to n$ processes whenever possible to speed up the calculation further. For non-MHV configurations, we use matrix elements from \mg~4~\cite{Alwall:2007st}. (Note that the use of \mg~4 puts some limitations on the configurations for which the relevant
information for MEC factors can be extracted easily from the 
matrix elements. In particular, this is the case for amplitudes 
with multiple quark pairs. These limitations will be lifted by a new interface to \mg~5 which is currently under development~\cite{mg5dev}.)

This article is organised as follows.
In \secRef{sec:shower}, we give an overview over the helicity-dependent
shower in \vc, including the extension to initial-state radiation 
The matrix-element correction formalism is reviewed in short in 
\secRef{sec:mecs} together with a brief introduction to the MHV
amplitudes in \vc. 
Results are presented in 
\secRef{sec:results}, before giving some concluding
remarks in \secRef{sec:conclusions}. The helicity-dependent antenna
functions are given in \appRef{app:helAnts}. \AppRef{app:implementation} summarises a few changes in the \vc code which we deem relevant to ensure that results obtained with the new implementation may be interpreted correctly, in particular in comparison with results obtained with earlier versions.

\section{Helicity-Dependent Showers}
\label{sec:shower}

A helicity-dependent antenna shower for final-state radiation has already been introduced 
in~\cite{Larkoski:2013yi}. The extension to hadronic initial states is straightforward. 
We start with a brief review of how emissions are generated and helicities  selected.
In cases where an event with unpolarised partons is showered by \vc, a polariser function is first called, which uses helicity matrix elements to assign explicit helicities to all partons. 
Since the events are also assigned colour flows, we first define the joint probability to select a parton configuration with a colour flow $i$ and a set of helicities $h$, 
\begin{equation} \label{eq:CheckForChosenHelicity}
P(h,i) = 
\underbrace{\frac{\mathrm{FC}^h}{\sum_{h'} \mathrm{FC}^{h'}}}_{\mbox{\scriptsize Helicity-Selection Factor}}  \times 
\underbrace{\frac{\mathrm{LC}_i^h}{\sum_{j} \mathrm{LC}^{h}_j}}_{\mbox{\scriptsize Colour-Flow Selection Factor}}  ~,
\end{equation}
where the full-colour (FC) and leading-colour (LC) matrix elements squared are defined by
\begin{eqnarray}
\mbox{FC}^h & = & \sum_{i,j} {\cal M}_i^h {\cal M}_{j}^{h*}~\label{eq:FC}\\
\mbox{LC}^h_i & = & |{\cal M}_i^h|^2
\end{eqnarray}
with ${\cal M}_i$ the amplitude for colour-ordering $i$. We also make use of the notation 
\begin{equation}
\mathrm{VC}_i^h  =  \mathrm{FC}^h \frac{\mathrm{LC}^h_i}{\sum_j \mathrm{LC}^h_j} \label{eq:VC}
\end{equation}
for the fraction of the full-colour helicity matrix element squared that is projected onto LC colour flow $i$.

As written here, the easiest would be to start by generating a helicity configuration, using the first factor in 
\eqRef{eq:CheckForChosenHelicity} and then subsequently generate a colour flow using the second factor. For events which already have colour-flow assignments, the conditional probability for  choosing helicity configuration $h$ is simplest to define in terms of the redefined LC matrix elements, 
\begin{equation} \label{eq:CondProb}
P(h\vert i) =
\frac{\mathrm{VC}^h_i}{\sum_{h'}\mathrm{VC}^{h'}_i}~.
\end{equation}
(If the corresponding matrix elements do not exist in \vc, the event will remain 
unpolarised and showered using helicity-averaged and  -summed antenna functions.) 

For events with explicit helicities, trial branchings are generated just as in the helicity-independent shower, i.e., using \emph{unpolarised} trial-antenna function overestimates. After generating 
the post-branching kinematics (see, e.g.,~\cite{Giele:2007di,Fischer:2016vfv}), the total probability for accepting a branching (denoting  pre-branching partons by $AB$ and  post-branching ones by $ijk$)\footnote{This is the same labelling convention as used in the \vc reference for final-state helicity showers~\cite{Larkoski:2013yi}.} is:
\begin{align}
P_\text{accept}=\frac{\antIdx{\text{phys}}}{\antIdx{\text{trial}}}=
\frac{\sum\limits_{h_{i},h_{j},h_{k}}\ant{h_{A},h_{B};h_{i},h_{j},h_{k}}}
{\antIdx{\text{trial}}}~,
\label{eq:PaccHel}
\end{align}
for fixed helicities $h_{A,B}$ of the parent partons. The sum over daughter helicities, $h_{i,j,k}$, in the physical antenna function, ${\cal A}_\mathrm{phys}$, runs over all possible (physical) helicities for the $ijk$ partons, with each term, $\ant{h_{A},h_{B};h_{i},h_{j},h_{k}}$, being a  
helicity-dependent antenna function. To avoid clutter, and for ease of reference, we  collect the precise forms for these functions in the appendix. We note that some of the functions differ (by nonsingular terms) from those used in previous versions of \vc, in particular those in \cite{Larkoski:2013yi,Fischer:2016vfv}.
We also note that the accept probability defined by \eqRef{eq:PaccHel} is in general identical to the unpolarised one (i.e., where one averages over $h_{A}$ and $h_{B}$ as well), up to  nonsingular terms. In case of initial-state radiation, \eqRef{eq:PaccHel} will be multiplied 
with the accept probability for the PDF ratios, just as in the unpolarised case~\cite{Fischer:2016vfv}. 

Explicit helicities are then selected for the daughters according to the relative probabilities given by the 
antenna functions,
\begin{align}
P(h_{A},h_{B};h_{i},h_{j},h_{k})=\frac{\ant{h_{A},h_{B};h_{i},h_{j},h_{k}}}
{\sum\limits_{h_{i},h_{j},h_{k}}\ant{h_{A},h_{B};h_{i},h_{j},h_{k}}}~,
\end{align}
where the denominator is equal to the numerator in \eqRef{eq:PaccHel}. Helicities are assigned to initial-state partons as well, using the same formalism. With the assumption that positive-helicity  partons
appear equally often as negative-helicity ones in the (anti)proton, the algorithm does not require
any modifications when considering initial-state partons. 

Helicity conservation implies that, for gluon emission off (massless) quarks or final-state gluons, the parent partons do not change their helicities.  A subtlety arises, however, for emissions off initial-state gluons. In the perspective of forwards evolution, such a branching looks like $g_{i}^\mrm{\,I}\to g_{A}^\mrm{\,I}
g_{j}^\mrm{\,F}$, where superscript $I$ ($F$) denotes an initial-state (final-state) parton; clearly, the helicity of parton $i$ can be inherited by either 
parton $j$ or parton $A$ without violating helicity conservation. Hence the reader should not be confused by the appearance of physical initial-state antenna functions for which $h_A \ne h_i$ in \appsRef{app:QGemit} and 
\ref{app:GGemit}, with corresponding DGLAP limits given in \appRef{app:GemiOffISG}.

LO antenna functions such as the ones discussed here are of course only accurate in the single-unresolved soft and collinear limits.
To estimate the amount of uncertainty caused by physical shower emissions being away from these limits, we use a two-pronged approach based on reweighting~\cite{Giele:2011cb}: 1) variation of the \emph{nonsingular terms} of the antenna functions to estimate how close a given branching is to the logarithmically-dominated region, and 2) variations of the antenna-function \emph{renormalisation scale} to estimate the potential impact of subleading-logarithmic terms. 
We emphasise that both types of variations are performed so that they  preserve the total cross section (i.e., the variations appear with equal and opposite signs in real and virtual corrections, respectively~\cite{Mrenna:2016sih}). The technical implementation in \vc is quite similar to that in \py~8; see the respective HTML User Manuals and  \appRef{app:implementation}.
The variation of the renormalisation scale in a helicity-dependent shower 
is performed just as for an unpolarised shower,
\begin{align}
\alpha_s(t)~~\rightarrow~~\alpha_s(k\,t)~.
\end{align}
Optionally, an NLO-level compensating term can be introduced for gluon emission, which forces the variation to be equal to the result for $k=1$ through order $\alpha_s^2$:
\begin{align}
\alpha_s(t)~~\rightarrow~~\alpha_s(k\,t)\big[1+(1-z)~\alpha_s\left(
\mrm{max}\left(m_\mrm{ant},k\,t\right)\right)
\,b_0\,\ln(k)\big]~,
\end{align}
where $b_0=(33-2n_F)/6\pi$, $n_F$ is the number of active flavours at the scale
$t$ and $m_\mrm{ant}$ the mass of the parent antenna. The prefactor $z$ is $s_{ik}/m^2_\mrm{ant}$
for final-final and $m^2_\mrm{ant}/\mrm{max}(s_{ik},m^2_\mrm{ant}+s_{jk})$ for
initial-initial and initial-final branchings,
with the post-branching invariants $s_{ij}$, $s_{jk}$, and $s_{ik}$,
The variation of the antenna functions by nonsingular terms,
\begin{align}
\ant{s_{ij},s_{jk},m^2_\mrm{ant}}~~\rightarrow~~
\ant{s_{ij},s_{jk},m^2_\mrm{ant}}+\frac{C_\mrm{NS}}{m^2_\mrm{ant}}~,
\end{align}
is performed such that the additional nonsingular term $C_\mrm{NS}/m^2_\mrm{ant}$
is distributed evenly amongst all helicity configurations for a specific antenna 
function, i.e.\ all helicity-dependent antenna functions obtain the same fraction 
of the nonsingular term. Note also that the nonsingular-term variations are cancelled
by ME corrections (up to the corrected order) and  therefore only need to be carried out for uncorrected orders. 

For any given (bin of a) physical observable, a large dependence on $C_\mrm{NS}$ indicates that corrections from hard matrix elements with higher numbers of legs are needed, while a significant dependence 
on the renormalisation scale indicates a need for further corrections at the loop level. 

Finally, it is worth emphasising that the statistical fluctuations of the uncertainty variations are generally larger than for the central (non-varied) predictions. This is due to the central prediction being unweighted (in our setup) and the the variations being computed by reweighting. See \cite{Bellm:2016voq} for an example of how  weighting (``biasing'') the central distribution can improve the relative statistical precision of the uncertainty bands.

\section{Matrix-Element Corrections and MHV amplitudes in \vc}
\label{sec:mecs}

\subsection{Matrix-Element Corrections}
The GKS formalism for iterated matrix-element corrections~\cite{Giele:2011cb} was originally based on so-called smoothly ordered showers, with a Markovian (history-independent) choice of restart scale after each branching. This allows the shower algorithm to generate phase-space points that violate the nominal ordering condition of the shower, at a suppressed but still non-zero rate, thus filling previously inaccessible regions of phase space; the correct (tree-level) emission rates can then be obtained via matrix-element corrections just as in the ordered part of phase space. However, general arguments indicate that the effective Sudakov factors for the non-ordered histories, are probably not correct~\cite{Hartgring:2013jma,Fischer:2016vfv,Li:2016yez}. Recent efforts~\cite{Li:2016yez,Fischer:2017yja} have therefore shifted focus back to filling the phase space for multiple hard emissions while remaining within the paradigm of strong ordering. In particular, we take the strongly-ordered iterated-MEC formalism presented in \cite{Fischer:2017yja} as our starting point, and adapt it to include explicit helicities. 

The question of Markovian vs non-Markovian behaviour comes about since the value of the shower evolution parameter in conventional strongly-ordered showers depends on which parton was the last one to be emitted. This cannot be uniquely determined merely by considering a given parton configuration; the value is a function of what shower history (or path) led to the configuration in question; a non-Markovian aspect. In the context of iterated ME corrections, non-Markovianity implies that the MEC factors contain nested  sums over shower histories involving clusterings all the way back to the Born configuration (while a Markovian algorithm only requires a single level of clusterings~\cite{Giele:2011cb}). 

Within the formalism presented in \cite{Fischer:2017yja}, the splitting kernels are redefined by multiplying them with the correction 
factor
\begin{align}
\mc R(\Phis{n+1}) = \left| \mc M(\Phis{n+1}) \right|^2\Big[ &
\sum\limits_{\Phi_{n}'}\ant{\Phis{n+1}/\Phisp{n}}~\mc R(\Phisp{n})
\sum\limits_{\Phi_{n-1}'}\Theta(t(\Phisp{n}/\Phisp{n-1})-t(\Phis{n+1}/\Phisp{n}))~
\ant{\Phisp{n}/\Phisp{n-1}}~\mc R(\Phisp{n-1}) \nonumber \\
& \prod\limits_{k=n-2}^{k\le1}\left(\sum\limits_{\Phi_{k}'}
\Theta(t(\Phisp{k+1}/\Phisp{k})-t(\Phisp{k+2}/\Phisp{k+1}))~
\ant{\Phisp{k+1}/\Phisp{k}}~\mc R(\Phisp{k})\right)\nonumber \\
& \sum\limits_{\Phi_{0}'}\Theta(t(\Phisp{1}/\Phisp{0})-t(\Phisp{2}/\Phisp{1}))~
\ant{\Phisp{1}/\Phisp{0}}~\Theta(t(\Phisp{0})-t(\Phisp{1}/\Phisp{0}))~
\left|\mc M(\Phisp{0})\right|^2\Big]^{-1}~.
\label{eq:MECs}
\end{align}
$\left| \mc M(\Phi_{n+1}) \right|^2$ denotes the matrix element squared of the $\Phi_{n+1}$ 
state and $\ant{\Phi_{n+1}/\Phi_{n}'}$ the antenna function, associated with the clustering
$\Phi_{n+1}\to\Phi_{n}'$.
The denominator sums over all possible ways the shower could have produced the 
$n+1$-particle state $\Phi_{n+1}$ from a given Born state $\Phi_0'$, including the 
correct weights of every shower step on the way.
This yields the recursive structure of \eqRef{eq:MECs} and the dependence on the 
correction factors of the previous orders.
In addition the (process-dependent) scale $t(\Phisp{0})$, at which the shower starts the
evolution off the Born state is taken into account.

For a helicity-dependent correction, we modify \eqRef{eq:MECs}  such that, for a given
polarised $\Phi_n$ state, the sums over the intermediate states $\Phi_{n-1}\ldots\Phi_0$
are extended to include all possible helicity configurations. As an example, consider
a possible clustering of a final-state $q\bar q$ pair to a gluon. In the unpolarised
case, one term corresponding to the clustering $q\bar q\to g$ contributes with the 
respective unpolarised antenna function and matrix element (which both implicitly involve helicity sums of course). For a polarised $q_+\bar q_-$ pair,
two different clustered helicity states are possible, $q_+\bar q_-\to g_+$ and
$q_+\bar q_-\to g_-$, each contributing according to their antenna function and
matrix element. The evolution variable, however, is the same as in the unpolarised 
case. This concludes our discussion of helicity-dependent matrix element corrections. 

\subsection{MHV Basics} \label{Sec:MHV Basics}
For fast evaluation of certain types of helicity configurations \vc uses Maximally Helicity Violating (MHV) amplitudes. MHV amplitudes have the advantage of being compact analytical expressions which are independent of Feynman diagrams; see~\cite{Mangano:1990by, Dixon:1996wi} for reviews. In this section, we briefly introduce the concepts and notation relevant to understanding the conventions and properties of the small library of MHV amplitudes implemented in \vc. 

In the following we consider all particles to be outgoing and massless. We recall that in this limit a particle's helicity corresponds to its chirality, and define our spinors in the helicity basis:
\begin{equation}
v_{\mp}(p) = u_{\pm}(p) = \frac{1}{2}\left(1 \pm \gamma^5\right)u(p)~, \quad \quad \overline{v}_{\mp}(p) = \overline{u}_{\pm}(p) = \overline{u}(p)\frac{1}{2}\left(1 \mp \gamma^5\right)~.
\end{equation}
The notation $\langle i j\rangle$ and  $[i j]$ is used for inner products of such spinors:
\begin{align}
\overline{u}_-(i)u_+(j) \equiv& \langle i j \rangle = \sqrt{p_j^+} e^{i\phi_i} - \sqrt{p_i^+} e^{i\phi_j}~, \\ 
\overline{u}_+(i)u_-(j) \equiv& [ij] = \langle j i \rangle^*~,
\end{align}
in terms of the (light-cone) momentum $p_i^+ = p_i^0 + p_i^3$ and $e^{i \phi_i} = (p_i^1 + i p_i^2)/\sqrt{p_i^+}$. For more details about spinor inner products and their properties see e.g.\ \cite{Mangano:1990by, Dixon:1996wi}. Note that in recent literature one often finds the convention $[ij] = \langle i j \rangle^*$, which is different to above (see e.g.\ \cite{Dixon:2013uaa}).

In the all-outgoing convention, helicity conservation implies that at least two pairs of opposite-helicity partons must exist for an $n$-parton amplitude to be nonzero\footnote{E.g., think of $++ \to ++$ and cross the two incoming positive helicities to be outgoing negative ones.}. If the remaining $n-4$ partons are all chosen to be of the same helicity ($+$ or $-$), the amplitude is called maximally helicity violating (MHV), and has a remarkably simple structure. The first MHV amplitude to be discovered was the all-gluon Parke-Taylor amplitude~\cite{Parke:1986gb}. In the following years this was extended to include one~\cite{Kunszt:1985mg, Mangano:1987kp} and two~\cite{Xu:1986xb, Gunion:1985bp, Gunion:1986zg} quark pairs, as well as to the case of a quark pair and a massive vector boson which decays leptonically~\cite{Berends:1987me, Mangano:1988kk}. 

\paragraph{All-Gluon Amplitudes:} To use these amplitudes we first note that the colour information can be factorised from the kinematics. In the $n$-point all-gluon case we use:
\begin{equation} \label{eq:All-Gluon Colour Structure}
{\cal M}_n (g_1, g_2, \dots , g_n) = g_s^{n-2} \sum_{\sigma \in S_n/Z_n} \text{Tr} ( t^{a_{\sigma(1)}} \dots t^{a_{\sigma(n)}}) A_n(\sigma(p_1^{h_1}), \dots, \sigma(p_n^{h_n})) ~,
\end{equation}
where $g_s$ is the strong coupling ($g_s^2 = 4 \pi \alpha_s$), the normalisation convention is  $t^a=\lambda^a\sqrt{2}$ with $\lambda^a$ being the generators of $SU(3)$, $p_i$ is the gluon momentum, $h_i$ the gluon helicity, $\text{Tr} ( t^{a_{\sigma(1)}} \dots t^{a_{\sigma(1)}})$ the colour factor and  $A_n(\sigma(p_1^{h_1}), \dots, \sigma(p_n^{h_n}))$ the kinematic part of the amplitude. The sum is over all non-cyclic permutations $\sigma$ of the particles. The Parke-Taylor amplitude then describes the kinematic part of \eqRef{eq:All-Gluon Colour Structure} and is given by:
\begin{equation}
A_n (i^-, j^-) = i \frac{\langle i j \rangle^4}{\langle 1 2 \rangle \langle 2 3 \rangle \dots \langle n 1 \rangle} ~,
\end{equation}
where gluons $i$ and $j$ have negative helicity, and all other particles have positive helicity. 

\paragraph{One Quark Pair:} If we add a $q\bar{q}$ pair we require that the quark and antiquark have opposite helicities (consistent with the gluon having spin 1), and use the following colour basis:
\begin{equation} \label{eq:Quark Anti-Quark Colour Structure}
{\cal M}_n (q, g_1, g_2, \dots , g_{n-2}, \bar{q}) = g_s^{n-2} \sum_{\sigma \in S_{n-2}} ( t^{a_{\sigma(1)}}, \dots t^{a_{\sigma(n-2)}})_{ij} A_n(q^{h_q}, \sigma(p_1^{h_1}), \dots, \sigma(p_{n-2}^{h_{n-2}}), \bar{q}^{h_{\bar{q}}} ) \ ,
\end{equation}
where $q$, $h_q$, and $i$ ($\bar{q}$, $h_{\bar{q}}$, and $j$) are respectively the quark (anti-quark) momentum, helicity, and colour index; and the sum is over all permutations of the gluons. If the quark and gluon $i$ each have negative helicity and all other particles positive helicity, then the kinematic amplitude is the given by:
\begin{equation} \label{eq:General q qBar MHV}
A_n(q^-, i^-, \bar{q}^+) = \frac{\langle q i \rangle^3 \langle \bar{q} i \rangle}{\langle \bar{q} q \rangle \langle q 1 \rangle \langle 1 2 \rangle \dots \langle (n-2) \bar{q} \rangle} ~,
\end{equation}
where the numbers refer to the (colour-ordered) gluons. If we exchange the helicities on the quarks, it is sufficient to exchange the exponents in the numerator:
\begin{equation}
A_n(q^+, i^-, \bar{q}^-) = \frac{\langle q i \rangle \langle \bar{q} i \rangle^3}{\langle \bar{q} q \rangle \langle q 1 \rangle \langle 1 2 \rangle \dots \langle (n-2) \bar{q} \rangle} ~.
\end{equation}

\paragraph{Two Quark Pairs:} The four-quark, $n-4$ gluon colour structure is given by:
\begin{align}
{\cal M}_n (q,&\bar{q},Q,\bar{Q},g_1, \dots,  g_{n-4}) = g_s^{n-2} \frac{A_0(h_q,h_Q,h_g)}{\lbrace q \bar{q} \rbrace \lbrace Q \bar{Q} \rbrace} \times  \nonumber \\
& \quad  \Big( \sum_{\sigma \in S_{n-4}}( t^{a_{\sigma(1)}} \dots t^{a_{\sigma(k)}})_{q\bar{Q}} ( t^{a_{\sigma(k+1)}} \dots t^{a_{\sigma(n-4)}})_{Q\bar{q}} \times A_n^{(0)}(q,1,\dots,k,\bar{q},Q,k+1,\dots,n-4,\bar{Q}) \nonumber \\
& \quad - \frac{1}{N_C} ( t^{a_{\sigma(1)}} \dots t^{a_{\sigma(k')}})_{q\bar{q}} ( t^{a_{\sigma(k'+1)}} \dots t^{a_{\sigma(n-4)}})_{Q\bar{Q}} \times A_n^{(1)}(q,1,\dots,k',\bar{q},Q,k'+1,\dots,n-4,\bar{Q}) \Big) ~, \label{eq:MHV Amplitude 4 Quarks}
\end{align}
where $\lbrace i j \rbrace = \langle i j \rangle$ for positive-helicity gluons and $\lbrace i j \rbrace = [j i]$ for negative-helicity gluons; $q$ and $Q$ label the two quark lines; $A_0(h_q,h_Q,h_g)$ is a kinematic function which depends on the helicities of the two quarks and the gluons, 
\begin{equation}
\begin{array}{c | c}
(h_q,h_Q,h_g) & A_0(h_q,h_Q,h_g) \\ \hline\\[-0.2cm]
(+,+,+) & \langle \bar{q}\bar{Q}\rangle^2 \\\\[-0.2cm]
(+,+,-) & [qQ]^2 \\\\[-0.2cm]
(+,-,+) & \langle \bar{q}Q\rangle^2 \\\\[-0.2cm]
(+,-,-) & [q\bar{Q}]^2 
\end{array}~,\label{eq:A_0 4 Quarks}
\end{equation}
with opposite-helicity cases obtained using parity transformation $\langle ij \rangle \leftrightarrow [ji]$; and the two functions  
$A_n^{(0)}$ and $A_n^{(1)}$ are kinematic amplitudes, for which we have used the short-hand notation $q \equiv q^{h_q}$, $i \equiv \sigma(p_i^{h_i})$ etc.: 
\begin{align}
A_n^{(0)} = & \frac{\lbrace q \bar{Q} \rbrace}{\lbrace q 1 \rbrace \lbrace 1 2 \rbrace \dots \lbrace k\bar{Q} \rbrace} \frac{\lbrace Q \bar{q} \rbrace}{\lbrace Q (k+1) \rbrace \lbrace (k+1) (k+2) \rbrace \dots \lbrace (n-4) \bar{q} \rbrace} ~, \\
A_n^{(1)} = & \frac{\lbrace q \bar{q} \rbrace}{\lbrace q 1 \rbrace \lbrace 1 2 \rbrace \dots \lbrace k\bar{q} \rbrace} \frac{\lbrace Q \bar{Q} \rbrace}{\lbrace Q (k+1) \rbrace \lbrace (k+1) (k+2) \rbrace \dots \lbrace (n-4) \bar{Q} \rbrace} ~. \label{eq:MHV 4 Quark Kinematic Amplitude}
\end{align}
We must sum over all possible partitions of gluons between the two quark colour lines, and also over all possible permutations of gluons within those partitions. If there are no gluons propagating off a particular colour line, then that colour line is described by a Kronecker delta. Note that this decomposition only works for the MHV configuration.

\paragraph{Drell-Yan, DIS, and hadronic Z decays:} To create MHV amplitudes with a single quark pair, a single lepton pair, and an arbitrary number of gluons, the four-quark amplitude can be recycled with all gluons coming from a single quark line. The second quark line is now equivalent to a $l \bar{l}$ pair up to couplings and an overall propagator factor. The amplitude then has the form
\begin{equation}
{\cal M}_{n}(h_q,h_{l}, h_g) = ig_s^{n-4}  \sum_{\sigma \in S_{n-4}} ( t^{a_{\sigma(1)}} \dots t^{a_{\sigma(n-4)}})_{ij} A_n(q^{h_q},\sigma(p_1^{h_1}), \dots, \sigma(p_{n-4}^{h_{n-4}}), \bar{q}^{h_{\bar{q}}}, l^{h_l}, \bar{l}^{h_{\bar{l}}}) ~,
\end{equation} 
where the sum is again over all gluon permutations. The kinematic amplitude is given by
\begin{equation}
A_n(q,1,\dots,n-4,\bar{q},l,\bar{l}) = \sum\limits_{V = \gamma, Z, W^{\pm}} M_V^{l} (h_{l}, h_q, h_g) \frac{1}{\lbrace q 1\rbrace \lbrace 1 2 \rbrace \dots \lbrace (n-4) \bar{q} \rbrace} ~,
\end{equation}
where the braces have the same meaning as in \eqRef{eq:MHV 4 Quark Kinematic Amplitude}, and the function $M_V^l$ is given by
\begin{equation}
M_V^{l} (h_{l}, h_q, h_g) = \frac{A_0(h_{l}, h_q, h_g)[\bar{l} l] (v_{h_l}^l) _V (v_{h_q}^q) _V}{\langle l \bar{l} \rangle [\bar{l} l] - M_V^2 + i\Gamma_V M_V} ~,
\end{equation}
where $A_0(h_{l}, h_q, h_g)$ is given by \eqRef{eq:A_0 4 Quarks}, $(v_{h_l}^l)_V$ ($(v_{h_q}^q)_V$) is the coupling of lepton $l$ (quark $q$) with helicity $h_l$ ($h_q$) to vector $V$, and $M_V$ and $\Gamma_V$ are the mass and width of the vector boson respectively.

Finally, we remark that in all of the above expressions, flipping the helicity of every particle is equivalent to exchanging each $\langle i j \rangle \leftrightarrow [j i]$. This concludes our brief recapitulation of the basics of the MHV formalism and convention choices.
 
\subsection{MHV within \vc}
The MHV amplitudes that are made available in standalone \vc are summarised in
 \tabRef{Table:MHV processes}. Note that these amplitudes are so far only used for QCD $2\to n$ matrix-element corrections, and that the second quark pair must have a different flavour to the first. 
\begin{table} 
\begin{center}
\begin{tabular}{c | c}
Type of process & Number of particles\\ \hline
All-gluon & 4-6 \\ 
1 quark pair plus gluons & 4-7 \\ 
2 quark pairs plus gluons & 4,5 \\ 
1 lepton pair, 1 quark pair plus gluons & 4-9 \\
\end{tabular}
\caption{The types of processes available in \vc's MHV library.}
\label{Table:MHV processes}
\end{center}
\end{table}

The colour-summed squared matrix element is calculated using the following matrix equation:
\begin{equation}
\mathrm{FC} = \sum_{ij}A_{\sigma_i}^{\dagger} C_{ij} A_{\sigma_j} ~,
\end{equation}
where FC stands for the full colour-summed matrix element squared as in \eqRef{eq:FC}, $C_{ij}$ is a colour matrix obtained by multiplying the colour factor from permutation $\sigma_i$ onto the conjugate colour factor from $\sigma_j$, and the sum is over all colour orders. We optimise the all-gluon amplitudes by diagonalising $C_{ij}$ for the $4$ and $5$-gluon matrix elements, and partially diagonalising $C_{ij}$ for the $6$-gluon matrix element as done in~\cite{Mangano:1990by}. 

By default, \vc uses MHV amplitudes wherever possible to compute its matrix-element correction factors, thus ensuring the fastest possible run time. However, this can be turned off (e.g., for cross checks with amplitudes from \mg) using the flag \texttt{vincia:useMHVamplitudes}. 
To calculate an MHV ME correction, \vc actively crosses the initial-state partons into the final state, rearranges the partons into the correct colour order, calculates all of the explicit spinor products needed, and then calculates the matrix element squared. 

The calculation of ME corrections for MHV configurations exhibits the nice feature that all
clustered states in \eqRef{eq:MECs} are MHV configurations as well.
Helicity conservation does not allow $++\to -$ nor $--\to +$ clusterings (in the all-outgoing convention). 
This results in clustered states being
either MHV configurations themselves or unphysical states with a vanishing amplitude. Consider $n$ positive- and $2$ negative-helicity outgoing 
partons as an example. Here clustered states contain either $n-1$ positive- and $2$ negative-helicity partons (MHV) or $n$ positive- and 
$1$ negative-helicity partons (unphysical).

For instructions on how to use \vc for calculating spinor products or
MHV amplitude in a standalone context, see the online user 
guide~\cite{VinciaUserReference}.

\subsection{Polarising events with MHV}

The fact that \vc assigns helicities to unpolarised events, with relative probabilities according to the corresponding helicity matrix elements squared, was  briefly discussed in \secRef{sec:shower}. An interesting simplification occurs when all of the contributing amplitudes are of the MHV kind, as is, e.g., the case for all QCD $2\to 2$ and $2\to 3$ processes. The simplification follows by noting that the full-colour (FC) MHV matrix elements squared all have the following form (so long as there is at most one quark pair):
\begin{equation} \label{eq:FCmhvStruct2}
\mathrm{FC}^h = |A_n^h(1, \dots, n)|^2 \left|\sum_{\sigma} \frac{1}{\langle \sigma(1) \sigma(2) \rangle \dots \langle \sigma(n) \sigma(1) \rangle} \text{CF}(\sigma(1) \dots \sigma(n))\right|^2 \equiv M_n^h \left| \sum_{\sigma} F(\sigma) \right|^2~,
\end{equation}
 where $h$ is a label denoting the helicity assignments, $M_n^h \equiv |A_n^h(1, \dots, n)|^2$ is some function of the helicities and momenta, $\sigma$ is the relevant set of permutations, CF is the relevant colour factor at the amplitude level, and $\left| \sum_{\sigma} F(\sigma) \right|^2$ is the square of the sum over colour permutations. For example, in the all-gluon amplitude $A_n^h(1, \dots, n)$ could be $\langle i j \rangle^4$. We have therefore factored out the helicity information $M_n^h$  from the colour information. This also works for the LC matrix elements $\mathrm{LC}^h_i$ which are given by \eqRef{eq:FCmhvStruct2} above without the sum of permutations. That is, $\mathrm{LC}^h_i = M_n^h \left| F(\sigma_i) \right|^2$. Recall that the conditional probability defined in \eqref{eq:CondProb} used to pick helicities for configurations that already have colour assignments has the form:
\begin{equation}
P(h\vert i) = \frac{\mathrm{VC}^h_i}{\sum_{h'}\mathrm{VC}^{h'}_i} = \frac{\mathrm{FC}^h \mathrm{LC}^h_i}{ \sum_j \mathrm{LC}_j^h} \left[
\sum_{h'} \frac{ \mathrm{FC}^{h'} \mathrm{LC}_i^{h'} }{  \sum_k  \mathrm{LC}_k^{h'}  } 
\right]^{-1}~.
\end{equation}
We can use \eqRef{eq:FCmhvStruct2} to simplify this:
\begin{align}
P(h\vert i) & = \frac{M_n^h  \left| \sum_{\sigma} F(\sigma) \right|^2 M_n^h \left| F(\sigma_i) \right|^2}{ \sum_j M_n^{h} \left| F(\sigma_j) \right|^2 } \left[ \sum_{h'}\frac{ M_n^{h'}  \left| \sum_{\sigma'} F(\sigma') \right|^2 M_n^{h'} \left| F(\sigma_i) \right|^2 }{  \sum_k  M_n^{h'} \left| F(\sigma_k) \right|^2} \right]^{-1} \nonumber \\
& = M_n^h  \frac{\left| \sum_{\sigma} F(\sigma) \right|^2 \left| F(\sigma_i) \right|^2}{\sum_j \left| F(\sigma_j) \right|^2} \left[ \frac{\left| \sum_{\sigma'} F(\sigma') \right|^2 \left| F(\sigma_i) \right|^2 }{ \sum_k \left| F(\sigma_k) \right|^2 } \sum_{h'} M_n^{h'} \right]^{-1} \nonumber \\ 
& = 
\frac{ M_n^h}{\sum_{h'} M_n^{h'}}~.
\end{align}
This shows that our factorisation allows to use the much simpler expressions $M_n^h \equiv |A_n^h(1, \dots, n)|^2$ to polarise the process. QCD processes are non-chiral, so we explicitly calculate only half of the factors $M_n^h$ to polarise them, since the other half are equal by parity. For the mostly-plus helicity case the factors $A_n^h(1, \dots n)$ are
\begin{equation}
\begin{array}{c | c | c}
\text{Process} & \text{Negative-helicity particles} & A_n^h(1, \dots, n) \\ \hline 
 \text{All-gluon} & i, j & \langle i j \rangle^4 \\
\text{Single Quark Pair} & q, i & \langle q i \rangle^3 \langle \bar{q} i \rangle \\
\text{Single Quark Pair} & \bar{q}, i & \langle q i \rangle \langle \bar{q} i \rangle^3 \\
\text{Quark Pair and Lepton Pair} & -  & A_0(h_l, h_q, +) (v_{h_l}^l) _V (v_{h_q}^q) _V
\end{array}~,\label{eq:A_n^h}
\end{equation}
while the mostly plus factors are given by the usual parity relation. 

Note that this also holds for the full-colour amplitudes used for selecting helicities at the colour-summed level, cf.~\eqRef{eq:CheckForChosenHelicity},
\begin{equation}
P(h) = \frac{\mathrm{FC}^h}{\sum_{h'} \mathrm{FC}^{h'}} 
= \frac{  M_n^h}{\sum_{h'} M_n^{h'}} \frac{ \left| \sum_{\sigma} F(\sigma) \right|^2}{ \left| \sum_{\sigma'} F(\sigma') \right|^2} =  \frac{ M_n^h}{\sum_{h'} M_n^{h'}}~.
\end{equation}

The preceding argument also works for 4-quark MHV amplitudes with distinct quark pairs provided one changes \eqRef{eq:FCmhvStruct2} to include the second colour connection. However, this doesn't work for all 4-quark MHV amplitudes because there is an extra colour-connection when two identical quarks have the same helicity. Hence the colour factor depends on the helicity and cannot be factorised.

\subsection{Speed Comparisons}

At the level of a pure shower (before ME corrections are imposed), the change from helicity-summed to helicity-sampling radiation functions requires the generation of one more random number per $n \to n+1$ branching, to select the helicity of the emitted parton. This comes in addition to at least three random numbers for the one-particle phase space. All else being equal, a helicity-sampling shower should therefore not be more than a factor 4/3 slower than a helicity-summed one. (Similar arguments hold for the initial polarisation step for the hard process). However, since there are many common components which must be computed regardless of the choice of helicity treatment, one expects the effective slowdown of the full shower algorithm to be milder than this upper limit. This is also borne out by explicit tests with \vc, which exhibit slowdowns of less than 10\% when switching on helicity-sampling. (See also the first bin of \figRef{fig:speed} below.)

As a measure of the relative speed of  helicity-dependent vs helicity-summed ME corrections, and the difference between using MHV matrix elements or \mg~4 ones, we consider the following specific (but fairly representative) benchmark case: $qg\to qg$ Born-level processes, with a minimum $\hat{p}_\perp$ of 100 GeV, in $pp$ collisions with $E_\mathrm{cm} =  10~\mathrm{TeV}$. A technical point is that, for this comparison, we switch $g\to q\bar{q}$ branchings off in the shower, so that the generated shower configurations are all of the simple  $qg\to qg + \mathrm{gluons}$ type. This allows us to illustrate speeds of ME corrections with up to three additional legs while, if $g\to q\bar{q}$ branchings had been switched on, the current version of \vc is restricted to ME corrections with up to two additional legs. (This restriction will be lifted in a future update.)

\begin{figure}[t]\centering
  \includegraphics[width=0.6\textwidth]{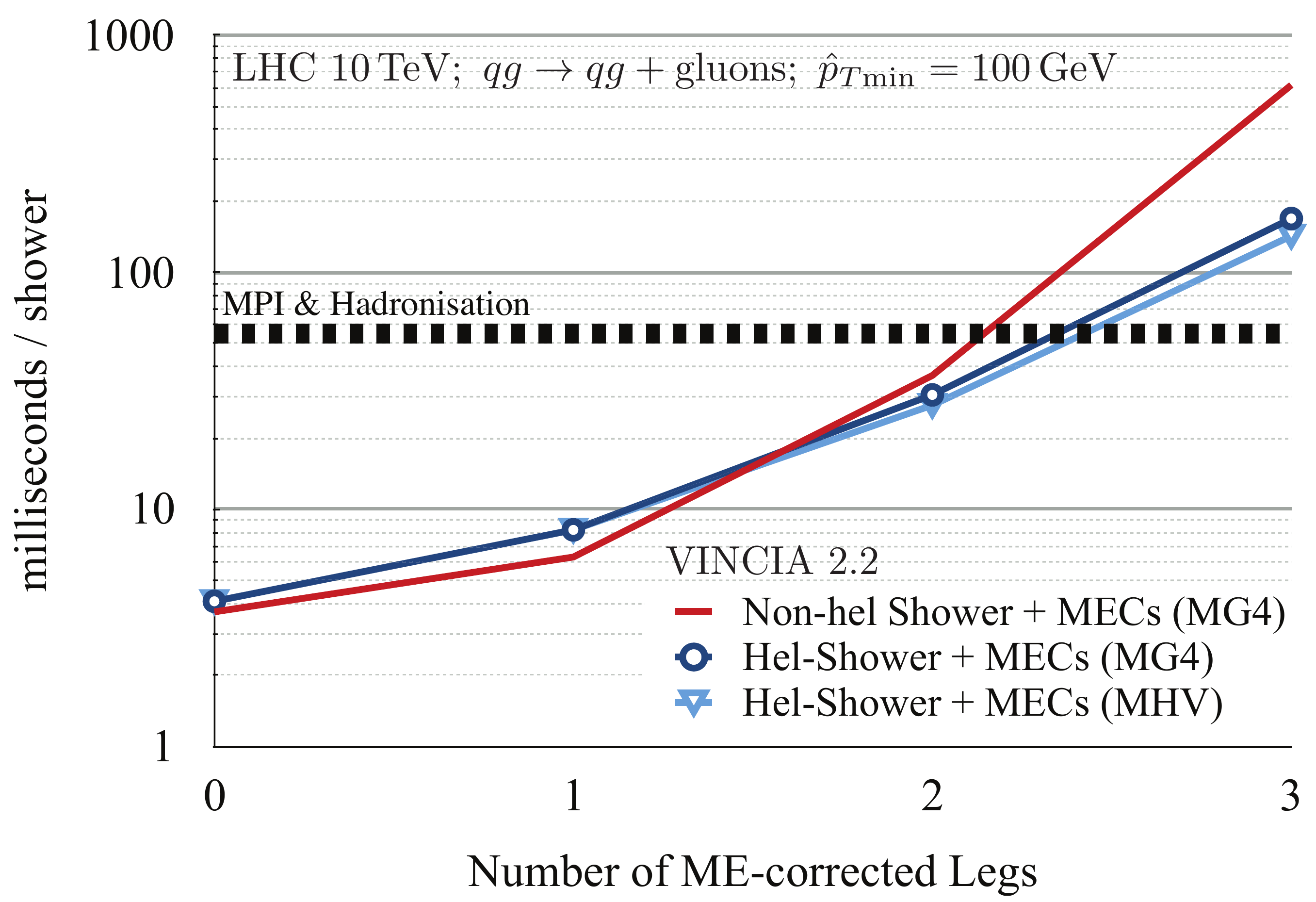}
  \caption{Speed comparison for  helicity-independent (``Non-hel'') and helicity-dependent (``Hel'') showers as a function of the number of ME-corrected legs, for $qg\to qg + \mathrm{gluons}$ with $\hat{p}_{\perp\mathrm{min}} = 100\,\mathrm{GeV}$, for $pp$ collisions at $E_{\mathrm{cm}} = 10\,\mathrm{TeV}$.  The dashed horisontal line indicates the time it takes to generate MPI and hadronisation for the same events. 
Results were obtained from 10,000 events generated for each run, on a single 2.9 GHz Intel Core i7 processor, using the \texttt{clang} compiler (v3.9), with -O2 optimisation.   \label{fig:speed}}
\end{figure}

\FigRef{fig:speed} illustrates  the number of milliseconds it takes to generate one shower, as a function of the number of legs that are requested to be ME-corrected. The solid (red) line without symbols uses helicity-summed showers and matrix elements, while the two blue curves (with symbols) show the dependence of the helicity-dependent formalism, with or without enabling the library of MHV matrix elements, respectively. For reference, the thick dashed horisontal line shows the time it takes to generate multi-parton interactions (MPI) and hadronisation for the same events\footnote{The thickness of the dashed line reflects that the helicity-dependent showers result in slightly longer MPI generation times due to the slightly slower showering off the MPI systems.}. For 0 or 1 corrected emissions, the helicity-summed shower is actually slightly faster, since the Born-level polariser and the helicity selection in the shower take a little extra time and the first-order ME corrections are very quick to evaluate even when summing over helicities. At two legs, however, the helicity-dependent formalism is up to 30\% quicker (with the MHV library  switched on) than the helicity-summed one. At three legs, the difference is a factor 4, with the MHV library allowing to shave an extra  $\sim 15\%$ off the shower-generation time relative to  using only MG4 matrix elements. 

One also notices that by two corrected legs, the showering time is becoming comparable to the time it takes to generate MPI and hadronisation for the events, hence this is the point at which the showering speed would start to be felt in the context of generating full events. By three corrected legs, the ME corrections dominate the event-generation time. 
The default in the current  version of \vc is  that ME corrections are enabled for QCD $2\to 2$ processes up to two additional legs; the event-generation time should therefore stay within roughly a factor 2 of that of the uncorrected algorithm. The complete set of matrix elements required for 3rd-order corrections will be provided in a future update. For hadronic $Z$, $W$, and $H$ production or decay, the full set of 3rd-order matrix elements are already available in the current version. (We note that the implementation of the iterated-MEC algorithm itself is general and could in principle handle any number of legs, if provided with the required matrix elements.)

\section{Example Application}
\label{sec:results}

To illustrate the properties of the ME-corrected algorithm (and uncertainty variations) in the context of a realistic application, we consider showers off $gg\to gg$ Born-level events and compare \py~8.226 and \vc~2.200 on three  observables sensitive to different aspects of the evolution: early branchings, late branchings, and a polarisation effect, respectively:
\begin{enumerate}
\item \textbf{Early branchings:}~the 3-jet resolution scale, $d_{23}$, using the longitudinally invariant $k_\perp$-jet algorithm with $R = 0.4$.  
The analysis is adapted from the code used in~\cite{Hoche:2015sya}, 
originally written by S.~H{\"o}che in the \riv~\cite{Buckley:2010ar} analysis
framework.
\item \textbf{Late branchings:}~the 6-jet $k_\perp$ resolution scale, $d_{56}$, with the same jet algorithm and analysis as above.
\item \textbf{Gluon polarisation:}~the angle between the event plane (characteristic of the original $gg\to gg$ Born-level event) and the plane of a subsequent $g\to b\bar{b}$ splitting. Here, the anti-$k_\perp$ jet algorithm with $R=0.2$ is used (so that the $b$ jets can be resolved down to small separations), and we impose a minimum jet $p_\perp$ of $50\,\mrm{GeV}$. The analysis is performed in the \riv~+~\fj~\cite{Cacciari:2011ma} framework. (For further ideas on how to exploit heavy-flavour tags to probe $g\to q\bar{q}$ splittings at colliders, see e.g.~\cite{Nachman:2017mmo,Ilten:2017rbd}.)
\end{enumerate}
The basic $2\to2$ QCD process is sampled with the cut $\hat p_\perp\ge 500\,
\mrm{GeV}$ on the final-state partons.
For consistency with the shower $\alpha_s$ parameters, \vc's default tune uses 
two-loop running for the strong coupling with $\alpha_s(m_Z^2)=0.118$ for the 
hard process. To compare predictions on an equal footing we apply the same
settings for the underlying Born process in \py.
To focus on the showering off the hard process all comparisons are done with multiparton interactions switched off. 

\begin{figure}[t]
\centering
\begin{minipage}[b]{0.49\textwidth}
  \rivetFigureRatio{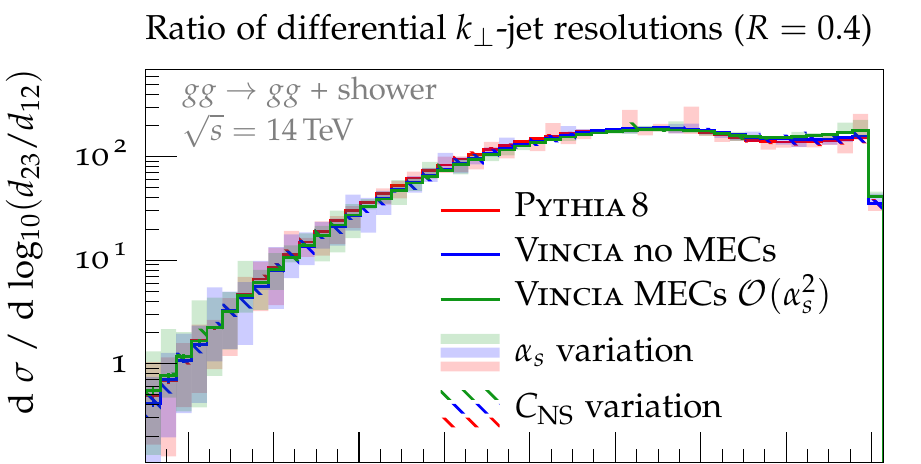}
  \rivetFigureRatio{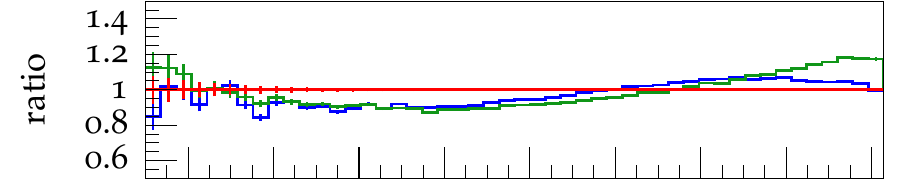}
  \rivetFigureRatio{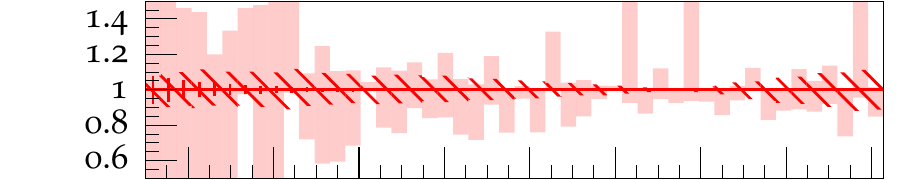}
  \rivetFigureRatio{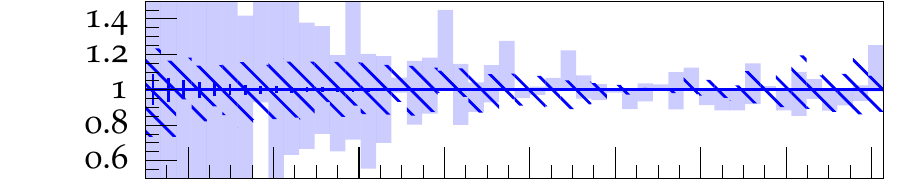}
  \rivetFigure{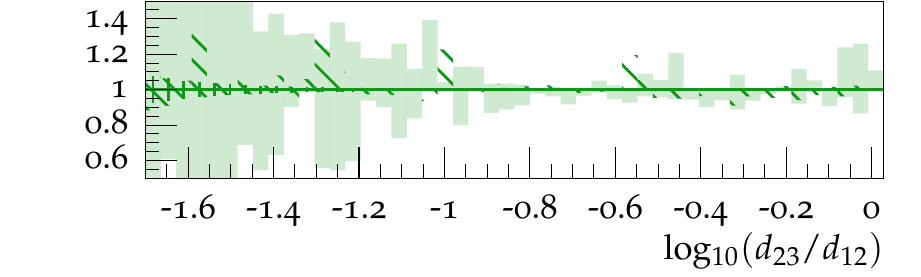}
\end{minipage}
\hfill
\begin{minipage}[b]{0.49\textwidth}
  \rivetFigureRatio{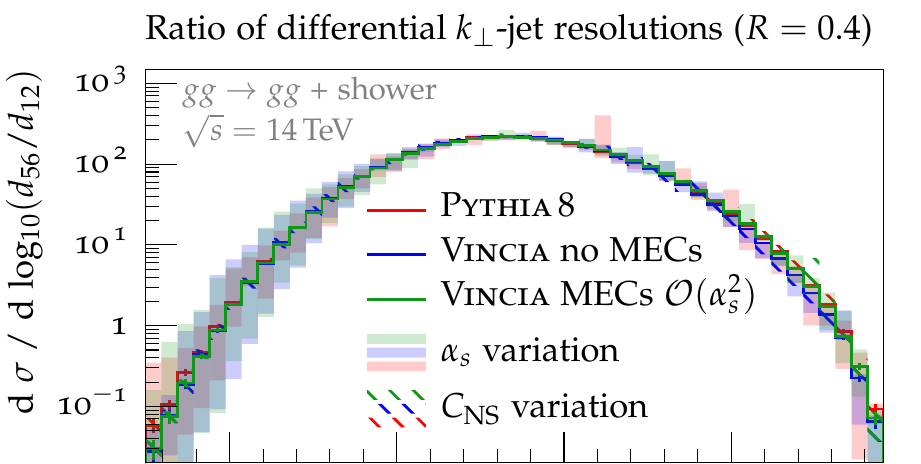}
  \rivetFigureRatio{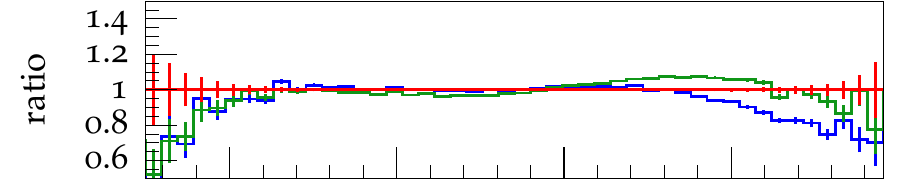}
  \rivetFigureRatio{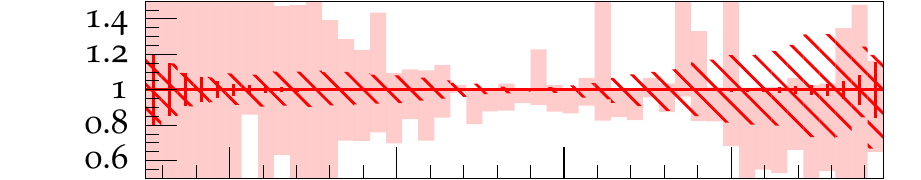}
  \rivetFigureRatio{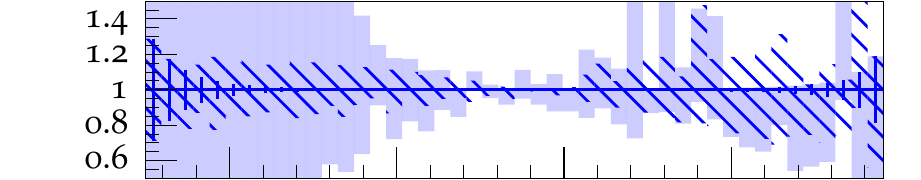}
  \rivetFigure{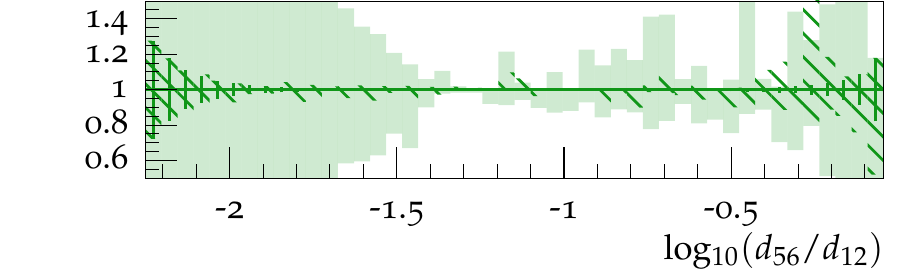}
\end{minipage}
\caption{Logarithmic distributions of ratios of differential jet resolutions,
$d_{23}/d_{12}$ and $d_{56}/d_{12}$, for showering $gg\to gg$ events. 
Predictions of \py8.226 and \vc2.200 with and without ME corrections are shown. 
The solid bands present a renormalisation-scale variation with $k=1/2$ and $2$ 
and the hashed bands a variation of the nonsingular terms with $C_\mrm{NS}=
\pm2$.
\label{fig:jetScaleRatios}}
\end{figure}

To obtain dimensionless variables, the jet resolution measures $d_{23}$ and 
$d_{56}$ are normalised by a factor $1/d_{12}$, i.e., they are effectively 
measured relative to a scale representing the $\hat{p}_\perp^2$ scale of the 
underlying Born process\footnote{This is similar to how, e.g., $m_Z^2$ is used 
to normalise corresponding observables in $e^+e^-$ collisions at the $Z$ pole.}.
The resulting quantities exhibit a
fixed-order behaviour for large values and a Sudakov suppression for low
values. Especially for well-resolved radiation, we therefore expect these
observables to be sensitive to low-order ME corrections, and hence the 
uncertainty associated with nonsingular-term variations should be reduced when 
\vc's ME corrections are switched on. (Note: \py does not incorporate ME 
corrections for QCD $2\to2$ processes.) Parton-level results for showered 
$gg\to gg$ events are presented in \figRef{fig:jetScaleRatios} with 
uncertainty bands.

The ME corrections in strongly-ordered events exhibit a modest effect of
up to $20\%$ for large values of $d_{23}/d_{12}$ and $d_{56}/d_{12}$, with 
the ME-corrected rate being larger than that of the pure \vc shower. 
Shape differences between the predictions of \py and \vc are visible throughout 
most of the distributions, with the uncorrected \vc shower generating a somewhat harder $d_{23}/d_{12}$ spectrum than \py. ME corrections increase the rate for large
$d_{56}/d_{12}$ values, bringing the predictions of \vc closer to that of \py.
Given the different choices 
of shower $\alpha_s$ parameters, evolution variable, and radiation functions, we do not consider this level of disagreement between the two models surprising. 
The evolution of the hard process starts at the factorisation scale 
for both showers. However, depending on the form of evolution variable,
the hardest possible scales correspond to different values of $d_{23}$.

All predictions exhibit some rather large fluctuations in the uncertainty
bands. The dijet system with the cut $\hat p_\perp\ge 500\,\mrm{GeV}$ as 
underlying hard process is typically accompanied
by a large number of additional jets. Given the nature of the reweighting
algorithm of~\cite{Mrenna:2016sih} (and similarly for \cite{Bellm:2016voq,
Bothmann:2016nao}) this may easily result in  fluctuating weights. In
addition we expect larger fluctuations in the nonsingular-term variations for 
the helicity shower, compared to the helicity-independent one. As discussed
in \secRef{sec:shower}, the additional nonsingular terms are
distributed evenly between all helicity configurations. This results in a
larger spread of weights, when considering helicity configurations that
constitute either a large or a small fraction of the helicity-summed
antenna functions. To mitigate the effects of weight fluctuations, we conclude that further development of these reweighting 
methods would be useful, in particular for large 
phase spaces (long shower chains). E.g., the authors in \cite{Bellm:2016voq} have demonstrated that combining biasing with reweighting can improve the relative statistical precision of the uncertainty variations, at the price of generating some reasonably well-behaved weights for the central (non-varied) event sample.

The variation of the nonsingular terms (hashed bands) results in a larger band 
around small  $|d_{23}/d_{12}|$ and $|d_{56}/d_{12}|$ for \vc without ME 
corrections, compared to \py.
The ME corrections cancel the effect of varying the nonsingular terms in the 
radiation functions. Consequently, the respective uncertainty band for \vc with 
ME corrections is very narrow, especially for $d_{23}$. The 
renormalisation-scale variations (shaded bands) are quite similar in size for 
all predictions. They show the largest effect for small jet separation scales,
where soft emissions and the Sudakov factor contribute to the distribution.

\begin{figure}[t]
\centering
\rivetFigure{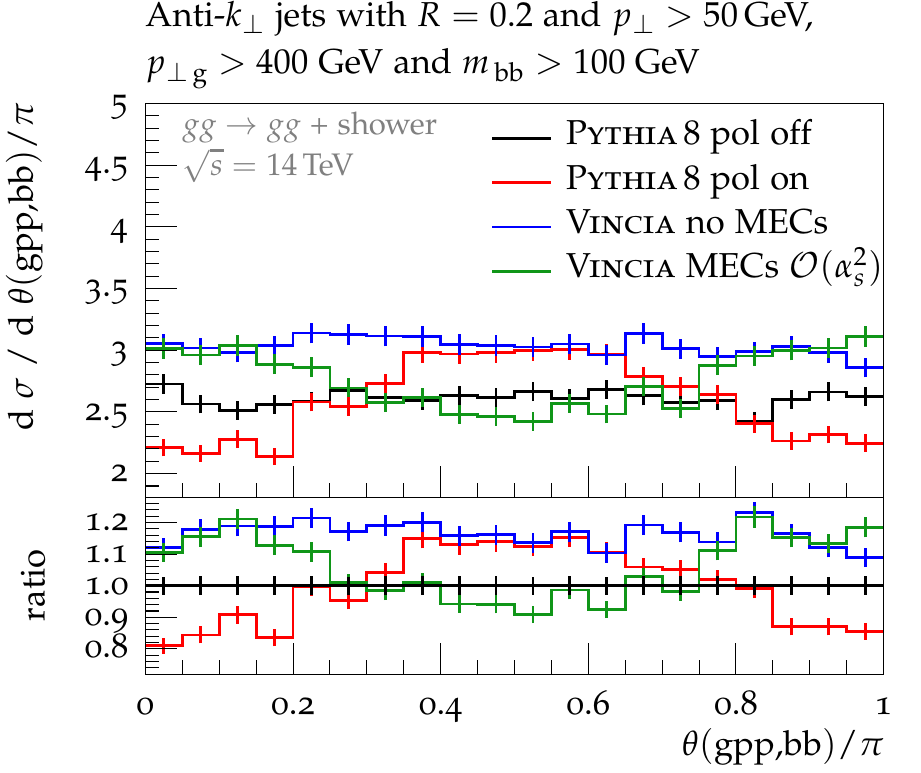}
\caption{The angle between the plane of the two $b$-jets and the plane of
the gluon jet and the beam axis. Predictions of \py8.2.26 and 
\vc2.2.0 with and without ME corrections are shown. In the labelling, 
``pol off'' refers to the \py8 parameters \ttt{TimeShower:phiPolAsym} and
\ttt{TimeShower:phiPolAsymHard} being switched off and ``pol on'' to the 
default settings, where both parameters are switched on.
\label{fig:angle}}
\end{figure}

We now turn to an observable where polarisation effects are expected to 
contribute. In events with two $b$-jets a plane is defined by the two jets. 
A second plane is defined by the gluon-jet (the sum of the two $b$-jets)
and the beam axis. In \figRef{fig:angle} the angle between the two planes is 
shown. A flat distribution is obtained with \py without gluon polarisation 
effects in the final-state shower and \vc without ME corrections. However,
\vc produces an around $15\%$ higher total rate, compared to \py. We note 
that both codes generate a similar total rate of $g\to b\bar b$ splittings in 
the shower, where the gluon splittings occur ``later'' in the evolution in
\py (i.e., preceded by a larger number of other branchings). 
The $b$-quarks are therefore more likely to obtain a smaller invariant
mass and might be clustered within the same jet. Together with the $p_\perp$
and invariant mass cuts on the jets, this may cause a smaller rate of events
with two $b$-jets.
The polarisation effects in \py leave the total rate unchanged, but increase the 
amount of events where the angle is close to $\pi/2$. The ME corrections in \vc
change the total rate by decreasing the number of events with splitting angles near 90 degrees. The qualitative effect is therefore the \emph{opposite} of that in \py, where the total shower rate is preserved, but the region around 90 degrees is enhanced by the polarisation effect. We conclude that a measurement of this observable, and the development of alternative strategies for corrections beyond fixed order (e.g., along the lines proposed in \cite{Li:2016yez}), would be desirable.

\section{Conclusions}
\label{sec:conclusions}

We have presented a helicity-dependent antenna shower for QCD initial- and final-state
radiation, implemented in the \vc shower model. The iterated ME correction formalism of \cite{Giele:2011cb,Larkoski:2013yi,Fischer:2016vfv, Fischer:2017yja} has been extended to cope
with helicity-dependent clusterings and splitting kernels involving initial-state legs, and in this work has been applied to strongly ordered showers in a direct extension of the formalism presented in~\cite{Fischer:2017yja}.  
We further reported on new, user-specifiable uncertainty variations in \vc, 
including renormalisation-scale and splitting-kernel variations. 

The new approach
and a library for tree-level MHV amplitudes enable a faster evaluation of MEC factors, as illustrated explicitly for the process $qg\to qg\,+$\,gluons. While the pure shower is slightly slower due to the additional step of helicity selection, the evaluation of ME corrections can be done significantly faster when only a single or a few helicity matrix elements need to be evaluated per trial branching, relative to when helicity-summed matrix elements are used.

To illustrate the effect of the iterated ME corrections and uncertainty 
variations within the helicity-dependent shower, we considered a few representative 
 observables, based on showered $gg\to gg$ Born-level events. As expected, ME corrections reduce the overall amount of variation considerably in regions of relatively hard emissions, where  process-dependent nonsingular terms (captured by the matrix elements)  dominate over the universal logarithmic terms (captured by the showers). In regions of large scale hierarchies, the uncertainty due to renormalisation-scale variations dominates and remains uncompensated by tree-level ME corrections. 

We also showed a more complex example, the angle between a Born-level $gg\to gg$ event plane and the plane of a subsequent $g\to b\bar{b}$ splitting. In \py, a general implementation of gluon polarisation effects implies an enhancement of such splittings at 90 degrees to the original event plane (while the total shower rate of $g\to b\bar{b}$ splittings is preserved); while in \vc, ME corrections dominantly act to suppress the overall rate of $g\to b\bar{b}$ splittings. Moreover, the suppression is most active for the most well-resolved branchings (at 90 degrees), leading to an opposite-sign effect than the one in \py. We conclude that there is  a complex interplay between the  rate and the angular dependence of these branchings, and intend to investigate this further in future studies.

\paragraph{Acknowledgments:} AL and PS acknowledge support from the Monash-Warwick Alliance Development Fund Project ``Collider Physics". 
PS is the recipient of an Australian Research Council Future Fellowship, FT130100744.

\appendix

\section{Helicity-Dependent Antenna Functions}
\label{app:helAnts}

\subsection{Notation and Conventions}
We use capital letters to denote partons in the pre-branching $n$-parton configuration and lower-case letters to denote partons in the post-branching $(n+1)-$parton configuration. Incoming partons are denoted $a$, $b$, while final-state partons are denoted $i$, $j$, $k$. Thus, for example, an initial-final antenna branching is written $AK \to ajk$. 

The scaled branching invariants for final-final antenna functions are
\begin{align}
y_{ij} = \frac{s_{ij}}{m_{IK}^2}~,~~
y_{jk} = \frac{s_{jk}}{m_{IK}^2}~,\text{ and }~
y_{ik} = \frac{s_{ik}}{m_{IK}^2}~,
\end{align}
and the energy fractions
\begin{align}
x_j = 1-\frac{1}{1-\mu_I^2}~y_{ik}~\text{ and }~
x_k = 1-\frac{1}{1-\mu_I^2}~y_{ij}~,
\end{align}
with $\mu_I=m_i/m_{IK}$. 
The scaled branching invariants for initial-final antenna functions are
\begin{align}
y_{aj} = \frac{s_{aj}}{m_{AK}^2+s_{jk}}~,~~
y_{jk} = \frac{s_{jk}}{m_{AK}^2+s_{jk}}~,\text{ and }~
y_{ak} = \frac{s_{ak}}{m_{AK}^2+s_{jk}}~,
\end{align}
and for initial-initial antenna functions
\begin{align}
y_{aj} = \frac{s_{aj}}{m_{AB}^2+s_{aj}+s_{jb}}~,~~
y_{jb} = \frac{s_{jb}}{m_{AB}^2+s_{aj}+s_{jb}}~,\text{ and }~
y_{AB} = \frac{m_{AB}^2}{m_{AB}^2+s_{aj}+s_{jb}}~.
\end{align}

Note that, for gluon-emission antennae involving massive parent quarks, a helicity-independent negative correction to the eikonal is added, with helicity-summed average:
\begin{equation}
\Delta a^\mathrm{eik}_{\mathrm{mass}} = - \frac{2m_{I}^2}{s_{ij}^2}  - \frac{2m_{K}^2}{s_{jk}^2}~.
\end{equation}

For gluon-splitting antennae $(Xg\to X\bar{q}_jq_k)$, the mass correction is positive:
\begin{equation}
\Delta a^\mathrm{split}_{\mathrm{mass}} =  \frac{m^2_{j}}{m_{jk}^4} ~.
\end{equation}

\subsection{$\mathbf{Q\bar{Q}}$ parents: Gluon Emission}
\label{app:QQemit}

The helicity averages for $q\bar{q}\to qg\bar{q}$ antennae are
\begin{align}
\mbox{\bf FF~:~~}a(q_I q_K \to q_i g_j q_k) & = \frac{1}{m_{IK}^2}\left[ \frac{2y_{ik}}{y_{ij} y_{jk}}
  + \frac{y_{jk}}{y_{ij}} + \frac{y_{ij}}{y_{jk}} + 1  \right] = \frac{1}{m_{IK}^2}\left[ \frac{(1-y_{ij})^2 + (1-y_{jk})^2}{y_{ij} y_{jk}} 
   + 1 \right]~, \\
  \mbox{\bf II~:~~}  a(\bar{q}_A q_B \to \bar{q}_a g_j q_b) & =  \frac{1}{s_{AB}}\left[ 
  \frac{2 y_{AB}}{y_{aj} y_{jb}} + \frac{y_{jb}}{y_{aj}} +
  \frac{y_{aj}}{y_{jb}} + 1 \right] =  \frac{1}{s_{AB}}\left[ \frac{(1-y_{aj})^2
  + (1-y_{jb})^2}{y_{aj} y_{jb}}+ 1 \right]~, \\
 \mbox{\bf IF~:~~}  a(q_A q_K \to q_a g_j q_k) & = \frac{1}{s_{AK}}\left[\frac{(1-y_{aj})^2
  + (1-y_{jk})^2}{y_{aj} y_{jk}} + \frac32 - \frac{y_{aj}^2}{2} - \frac{y_{jk}^2}{2}\right]~,
  \end{align}
where the slightly different nonsingular terms chosen for the IF case ensure positivity of in particular the $(++ \to +-+)$  antenna function over all of the IF phase space, while the nonsingular terms for the FF and II cases result from averaging over the corresponding helicity matrix elements for $Z$ and $H$ decays. 

The individual helicity contributions are:
\begin{align}
a(++\to +++) & = \frac{1}{m^2_{IK}} \left[\frac{1}{y_{ij} y_{jk}} \right] ~, \\
a(++\to +-+) & = \frac{1}{m^2_{IK}} \left[\frac{(1-y_{ij})^2
  + (1-y_{jk})^2 -1}{y_{ij} y_{jk}} + 2\right] ~, \\
\raisebox{3.5ex}{\mbox{\bf FF~:}~~~~~}
a(+-\to ++-) & = \frac{1}{m^2_{IK}} \left[\frac{(1-y_{ij})^2}{y_{ij} y_{jk}}\right] ~, \\
a(+-\to +--) & = \frac{1}{m^2_{IK}} \left[\frac{(1-y_{jk})^2}{y_{ij} y_{jk}}\right] ~.
\intertext{~}
a(++\to +++) & = \frac{1}{s_{AB}}\left[ \frac{1}{y_{aj} y_{jb}}\right]~, \\
a(++\to +-+) & = \frac{1}{s_{AB}}\left[ \frac{y_{AB}^2}{y_{aj} y_{jb}}\right]~, \\
\raisebox{3.5ex}{\mbox{\bf II~:}~~~~~}a(+-\to ++-) & = \frac{1}{s_{AB}}\left[ \frac{(1-y_{aj})^2}{y_{aj} y_{jb}} \right]~, \\
a(+-\to +--) & = \frac{1}{s_{AB}} \left[\frac{(1-y_{jb})^2}{y_{aj} y_{jb}}\right]~.
\intertext{~}
a(++\to +++) & = \frac{1}{s_{AK}} \left[\frac{1}{y_{aj} y_{jk}}\right] ~, \\
a(++\to +-+) & = \frac{1}{s_{AK}} \left[\frac{(1-y_{aj})^2 + (1-y_{jk})^2 -1}{y_{aj}
  y_{jk}}+3-y_{aj}^2-y_{jk}^2\right] ~, \\
\raisebox{3.5ex}{\mbox{\bf IF~:}~~~~~}a(+-\to ++-) & = \frac{1}{s_{AK}} \left[\frac{(1-y_{aj})^2}{y_{aj} y_{jk}}\right] ~, \\
a(+-\to +--) & = \frac{1}{s_{AK}} \left[\frac{(1-y_{jk})^2}{y_{aj} y_{jk}}\right] ~.
\end{align}

\subsection{$\mathbf{QG}$ parents: Gluon Emission}
\label{app:QGemit}

The helicity averages for $qg\to qgg$ antennae are
\begin{align}
\mbox{\bf FF~:~~} a(q_I g_K \to q_i g_j g_k) & = \frac{1}{m_{IK}^2}\left[ 
  \frac{2y_{ik}}{y_{ij} y_{jk}} 
  + \frac{y_{jk}}{y_{ij}} + \frac{y_{ij}(1-y_{ij})}{y_{jk}}
  + y_{ij} + \frac{y_{jk}}{2} \right] \nonumber \\
& = \frac{1}{m_{IK}^2}\left[
  \frac{(1-y_{ij})^3 + (1-y_{jk})^2}{y_{ij} y_{jk}} - \frac{2 \mu_I^2}{y_{ij}^2} 
  +\frac{y_{ik}-y_{ij}}{y_{jk}} + 1 + y_{ij} + \frac{y_{jk}}{2}\right]~, \\
\mbox{\bf II~:~~} a(q_A g_B \to q_ag_j g_b) & = \frac{1}{s_{AB}}\left[ \frac{(1-y_{aj})^3
  + (1-y_{jb})^2}{y_{aj}y_{jb}} +
  \frac{1+y_{aj}^3}{y_{jb}(1-y_{aj})} + 2 - y_{aj} - \frac{y_{jb}}{2}\right]~, \\
\mbox{\bf IF~:~~} a(q_A g_K \to q_a g_j g_k) & = \frac{1}{s_{AK}}\left[\frac{(1-y_{aj})^3
  + (1-y_{jk})^2}{y_{aj} y_{jk}} + \frac{1-2y_{aj}}{y_{jk}}
  + \frac32 + y_{aj} - \frac{y_{jk}}{2} - \frac{y_{aj}^2}{2} \right]~, \\
\mbox{\bf IF~:~~} a(g_A q_K \to g_a g_j q_k) & = \frac{1}{s_{AK}}\left[ \frac{(1-y_{jk})^3
  + (1-y_{aj})^2}{y_{aj}y_{jk}} + \frac{1+y_{jk}^3}{y_{aj}(y_{AK}+y_{aj})}
  + \frac32 - \frac{y_{jk}^2}{2} \right]~. \\
\end{align}
Note that for the initial-final case two antennae, $qg\to qgg$ and
$gq\to ggq$, exist.

The individual helicity contributions are:
\begin{align}
a(++\to +++) & = \frac{1}{m^2_{IK}}\left[\frac{1}{y_{ij} y_{jk}}
  + (1-\alpha)(1-y_{jk})\left( \frac{1-2y_{ij}-y_{jk}}{y_{jk}} \right)\right] ~, \\
a(++\to +-+) & = \frac{1}{m^2_{IK}} \left[\frac{(1-y_{ij})y_{ik}^2}{y_{ij} y_{jk}} \right]~, \\
\raisebox{3.5ex}{\mbox{\bf FF~:}~~~~~}
a(+-\to ++-) & = \frac{1}{m^2_{IK}} \left[\frac{(1-y_{ij})^3}{y_{ij} y_{jk}}\right] ~, \\
a(+-\to +--) & = \frac{1}{m^2_{IK}} \left[ \frac{(1-y_{jk})^2}{y_{ij} y_{jk}} +
  (1-\alpha)(1-y_{jk})\left(\frac{1-2y_{ij}-y_{jk}}{y_{jk}} \right)\right] ~.
\intertext{~}
a(++\to +++) & = \frac{1}{s_{AB}} \left[\frac{1}{y_{aj}y_{jb}}
  \frac{1-y_{jb}}{1-y_{aj}-y_{jb}}\right] = \frac{1}{s_{AB}} \left[\frac{1}{y_{aj}
  y_{jb}} +  \frac{1}{ y_{jb}y_{AB}} \right] \label{eq:QGemitIIhelFirst} \\
& \stackrel{\mbox{sing}}{~\to~} \frac{1}{s_{AB}} \left[\frac{1}{y_{aj}y_{jb}}
  + \frac{1}{ y_{jb}(1-y_{aj})} \right] ~, \\
a(++\to +-+) & = \frac{1}{s_{AB}} \frac{1}{y_{aj} y_{jb}} \frac{y_{AB}^3}{1-y_{jb}}
  \frac{y_{AB}^3}{s_{AB}} \left[ \frac{1}{y_{aj} y_{jb}} + \frac{1}{y_{aj}
  (1-y_{jb}) } \right] \\
& \stackrel{\mbox{sing}}{~\to~}  \frac{1}{s_{AB}} \left[ \frac{y_{AB}^3}{y_{aj}
  y_{jb}} + \frac{y_{AB}^2}{y_{aj}} \right] = \frac{1}{s_{AB}} \frac{(1-y_{aj})
  y_{AB}^2}{y_{aj}y_{jb}} ~, \\
a(+-\to ++-) & = \frac{1}{s_{AB}} \left[\frac{(1-y_{aj})^3}{y_{aj}
  y_{jb}} + \frac{1-y_{jb} - y_{aj}^2}{1-y_{jb}} \right] \\
& \stackrel{\mbox{sing}}{~\to~} \frac{1}{s_{AB}} \frac{(1-y_{aj})^3}{y_{aj} y_{jb}} ~, \\
\raisebox{3.5ex}{\mbox{\bf II~:}~~~~~}
a(+-\to +--) & = \frac{1}{s_{AB}} \frac{1}{y_{aj} y_{jb}} \frac{(1-y_{jb})^3}
  {1-y_{aj}-y_{jb}} = \frac{(1-y_{jb})^2}{s_{AB}} \left[\frac{1}{y_{aj} y_{jb}}
  + \frac{1}{y_{jb}}\frac{1}{1-y_{aj}-y_{jb}} \right] \\
& \stackrel{\mbox{sing}}{~\to~} \frac{1}{s_{AB}} \left[\frac{(1-y_{jb})^2}{y_{aj}y_{jb}}
  + \frac{1}{y_{jb}(1-y_{aj})} \right] ~, \\
a(++\to +--) & = \frac{1}{s_{AB}}\frac{y_{aj}^3}{y_{jb}(1-y_{jb})}
  \frac{1}{1-y_{aj}-y_{jb}} ~, \\
& \stackrel{\mbox{sing}}{~\to~} \frac{1}{s_{AB}}\frac{y_{aj}^3}{y_{jb}(1-y_{aj})} ~, \\
a(+-\to +++) & = a(++\to +--) ~. \label{eq:QGemitIIhelLast}
\intertext{~}
a(++\to +++) & = \frac{1}{s_{AK}} \left[\frac{1}{y_{aj} y_{jk}}
  + \frac{1-2y_{aj}}{y_{jk}} \right] ~, \\
a(++\to +-+) & = \frac{1}{s_{AK}} \left[\frac{(1-y_{aj})^3
  + (1-y_{jk})^2 - 1}{y_{aj} y_{jk}} + 3 - y_{aj}^2\right] ~, \\
\raisebox{3.5ex}{\mbox{\bf IF~:}~~~~~}
a(+-\to ++-) & = \frac{1}{s_{AK}} \left[\frac{(1-y_{aj})^3}{y_{aj} y_{jk}}\right] ~, \\
a(+-\to +--) & = \frac{1}{s_{AK}} \left[\frac{(1-y_{jk})^2}{y_{aj} y_{jk}}
  + \frac{1-2y_{aj}}{y_{jk}} + 2y_{aj} - y_{jk} \right] ~.
\intertext{~}
a(++\to +++) & = \frac{1}{s_{AK}} \left[ \frac{1}{y_{aj} y_{jk}} + \frac{1}{y_{aj} (y_{AK}
  + y_{aj})} \right] ~, \\
a(++\to +-+) & = \frac{1}{s_{AK}} \left[\frac{(1-y_{aj})^2 + (1-y_{jk})^3 -1}{y_{aj}y_{jk}}
  + 3 - y_{jk}^2\right] ~, \\
a(+-\to ++-) & = \frac{1}{s_{AK}} \left[ \frac{(1-y_{aj})^2}{y_{aj} y_{jk}}
  + \frac{1}{y_{aj} (y_{AK} + y_{aj})} \right] ~, \\
\raisebox{3.5ex}{\mbox{\bf IF~:}~~~~~}
a(+-\to +--) & = \frac{1}{s_{AK}}\left[\frac{(1-y_{jk})^3}{y_{aj} y_{jk}}\right] ~, \\
a(++\to --+) & = \frac{1}{s_{AK}}\frac{y_{jk}^3}{y_{aj} (y_{AK} + y_{aj})} ~, \\
a(+-\to ---) & = a(++\to --+) ~.
\end{align}

Note that for gluons in the initial-state an additional helicity 
configuration\,\footnote{Additional with respect to the final-state antenna 
functions.} arises where the final-state gluon inherits the helicity.

\subsection{$\mathbf{GG}$ parents: Gluon Emission}
\label{app:GGemit}

The helicity averages for $gg\to ggg$ antennae are
\begin{align}
\mbox{\bf FF~:~~} a(g_I g_K \to g_i g_j g_k) & = \frac{1}{m_{IK}^2}\left[ 
  \frac{2 y_{ik}}{y_{ij} y_{jk}} 
  + \frac{y_{jk}(1-y_{jk})}{y_{ij}} + \frac{y_{ij}(1-y_{ij})}{y_{jk}} 
  + \frac12y_{ij} + \frac12 y_{jk}\right] \nonumber \\
& = \frac{1}{m_{IK}^2}\left[ \frac{(1-y_{ij})^3 + (1-y_{jk})^3}{y_{ij} y_{jk}}
  + \frac{y_{ik}-y_{ij}}{y_{jk}}
  + \frac{y_{ik}-y_{jk}}{y_{ij}} + 2 + \frac12y_{ij} + \frac12 y_{jk}  \right]~, \\
\mbox{\bf II~:~~} a(g_A g_B \to g_ag_j g_b) & = \frac{1}{s_{AB}} \left[ \frac{(1-y_{aj})^3
  + (1-y_{jb})^3}{y_{aj}y_{jb}}  + \frac{1+y_{aj}^3}{y_{jb}(1-y_{aj})} 
  + \frac{1+y_{jb}^3}{y_{aj}(1-y_{jb})} + 3 - \frac{3y_{aj}}{2}\right. \nonumber \\
  & \qquad\qquad \left.- \frac{3y_{jb}}{2} \right]~, \\
\mbox{\bf IF~:~~} a(g_A g_K \to g_a g_j q_k) & = \frac{1}{s_{AK}} \left[ \frac{(1-y_{aj})^3
  + (1-y_{jk})^3}{y_{aj}y_{jk}} + \frac{1 + y_{jk}^3}{y_{aj} (y_{AK} + y_{aj})} 
  + \frac{1-2 y_{aj}}{y_{jk}} + 3 -2y_{jk} \right]~.
\end{align}

The individual helicity contributions are:
\begin{align}
a(++\to +++) & = \frac{1}{m^2_{IK}} \left[ \frac{1}{y_{ij} y_{jk}}
  + (1-\alpha)\left( (1-y_{ij})\frac{1-2y_{jk}-y_{ij}}{y_{ij}}
  + (1-y_{jk})\frac{1-2y_{ij}-y_{jk}}{y_{jk}}\right) \right]~, \\
a(++\to +-+) & = \frac{1}{m^2_{IK}} \left[\frac{y_{ik}^3}{y_{ij} y_{jk}}\right]~, \\
\raisebox{3.5ex}{\mbox{\bf FF~:}~~~~~}
a(+-\to ++-) & = \frac{1}{m^2_{IK}} \left[ \frac{ (1-y_{ij})^3}{y_{ij} y_{jk}}
  +(1-\alpha)(1-y_{ij})\frac{1-2y_{jk}}{y_{ij}}\right]~, \\
a(+-\to +--) & = \frac{1}{m^2_{IK}} \left[ \frac{ (1-y_{jk})^3}{y_{ij} y_{jk}} 
  + (1-\alpha)(1-y_{jk})\frac{1-2y_{ij}}{y_{jk}} \right]~.
\intertext{~}
a(++\to +++) & = \frac{1}{s_{AB}} \left[\frac{1}{y_{aj} y_{jb}} 
  + \frac{1}{y_{jb}(1-y_{aj})} + \frac{1}{y_{aj}(1-y_{jb})} \right] ~,
  \label{eq:GGemitIIhelFirst} \\     
a(++\to +-+) & = \frac{1}{s_{AB}}\frac{y_{AB}^3}{y_{aj} y_{jb}} ~, \\
a(+-\to ++-) & = \frac{1}{s_{AB}} \left[\frac{(1-y_{aj})^3}{y_{aj} y_{jb}}  
  + \frac{1}{y_{aj}(1-y_{jb})} \right] ~, \\
a(+-\to +--) & = \frac{1}{s_{AB}} \left[\frac{(1-y_{jb})^3}{y_{aj} y_{jb}}  
  + \frac{1}{y_{jb}(1-y_{aj})} \right] ~, \\
\raisebox{3.5ex}{\mbox{\bf II~:}~~~~~}
a(++\to +--) & = \frac{1}{s_{AB}} \frac{y_{aj}^3}{y_{jb}(1-y_{aj})} ~, \\
a(++\to --+) & = \frac{1}{s_{AB}} \frac{y_{jb}^3}{y_{aj}(1-y_{jb})} ~, \\
a(+-\to +++) & = a(++\to +--) ~, \\
a(+-\to ---) & = a(++\to --+) ~. \label{eq:GGemitIIhelLast}
\intertext{~}
a(++\to +++) & = \frac{1}{s_{AK}} \left[  \frac{1}{y_{aj} y_{jk}}
  + \frac{1-2y_{aj}}{y_{jk}} + \frac{1}{y_{aj} (y_{AK}+y_{aj})} \right] ~, \\
a(++\to +-+) & = \frac{1}{s_{AK}} \left[  \frac{(1-y_{aj})^3
  +(1-y_{jk})^3-1}{y_{aj} y_{jk}} + 6 - 3y_{aj} - 3y_{jk} + y_{aj}y_{jk} \right] ~,\\
a(+-\to ++-) & = \frac{1}{s_{AK}} \left[ \frac{(1-y_{aj})^3}{y_{aj} y_{jk}}
  + \frac{1}{y_{aj} (y_{AK}+y_{aj})} \right] ~,\\
\raisebox{3.5ex}{\mbox{\bf IF~:}~~~~~}
a(+-\to +--) & = \frac{1}{s_{AK}} \left[ \frac{(1-y_{jk})^3}{y_{aj} y_{jk}}
  + \frac{1-2y_{aj}}{y_{jk}} + 3y_{aj} - y_{jk} - y_{aj} y_{jk} \right] ~, \\
a(++\to --+) & = \frac{1}{s_{AK}}   \frac{y_{jk}^3}{y_{aj} (y_{AK} + y_{aj})} ~, \\
a(+-\to ---) & = a(++\to --+) ~.
\end{align}

Note that for gluons in the initial-state an additional helicity 
configuration\,\footnote{Additional with respect to the final-state antenna 
functions.} arises where the final-state gluon inherits the helicity.

\subsection{$\mathbf{G\to \bar{Q}Q}$ Splittings}
\label{app:split}

The helicity averages for $Xg\to X\bar qq$ antennae (final-state gluon
splitting) are
\begin{align}
\mbox{\bf FF~:~~} a(X_I g_K \to X_i \bar{q}_j q_k) & = \frac{1}{2m_{jk}^2} 
  \left[ (1-x_j)^2 + (1-x_k)^2 \right] 
  = \frac{1}{2m_{jk}^2} \left[ \frac{y_{ik}^2 + y_{ij}^2}{(1-\mu_I^2)^2}
  \right]~, \\
\mbox{\bf IF~:~~} a(X_A g_K \to X_a \bar{q}_j q_k) & = \frac{1}{2m^2_{jk}} 
  \left[ y_{ak}^2 + y_{aj}^2 \right]~.
\end{align}

The individual helicity contributions are:
\begin{align}
a(X+\to X-+) & = \frac{1}{2m^2_{jk}} \frac{y_{ik}^2}{(1-\mu_I^2)^2} ~=~
  \frac{(1-x_j)^2}{2m_{jk}^2}~, \\
\raisebox{3.5ex}{\mbox{\bf FF~:}~~~~~}
a(X+\to X+-) & = \frac{1}{2m^2_{jk}} \frac{y_{ij}^2}{(1-\mu_I^2)^2} ~=~
  \frac{(1-x_k)^2}{2m_{jk}^2} ~.
\intertext{~}
a(X+\to X-+) & = \frac{y_{ak}^2}{2m^2_{jk}} ~, \\
\raisebox{3.5ex}{\mbox{\bf IF~:}~~~~~}
a(X+\to X+-) & = \frac{y_{aj}^2}{2m^2_{jk}} ~.
\end{align}

The helicity averages for $qX\to g\bar qX$ antennae (quark backwards
evolving to a gluon) are
\begin{align}
\mbox{\bf II~:~~} a(q_A X_B \to g_a\bar{q}_j X_b) & = \frac{1}{s_{AB}}
  \frac{y_{AB}^2 + (1-y_{AB})^2}{y_{aj}}~, \\
\mbox{\bf IF~:~~} a(q_A X_K \to g_a\bar{q}_j X_k) & = \frac{1}{s_{AK}} 
  \frac{y_{AK}^2 + (1-y_{AK})^2}{y_{aj}}~.
\end{align}

The individual helicity contributions are:
\begin{align}
a(+X \to +-X) & = \frac{1}{s_{AB}}\frac{y_{AB}^2}{y_{aj}} ~, \\
a(+X \to --X) & = \frac{1}{s_{AB}}\frac{(1-y_{AB})^2}{y_{aj}} ~, \\
\raisebox{3.5ex}{\mbox{\bf II~:}~~~~~}
a(-X \to -+X) & = \frac{1}{s_{AB}}\frac{y_{AB}^2}{y_{aj}} ~, \\
a(-X \to ++X) & = \frac{1}{s_{AB}}\frac{(1-y_{AB})^2}{y_{aj}} ~.
\intertext{~}
a(+X \to +-X) & = \frac{1}{s_{AK}}\frac{y_{AK}^2}{y_{aj}} ~, \\
a(+X \to --X) & = \frac{1}{s_{AK}}\frac{(1-y_{AK})^2}{y_{aj}} ~, \\
\raisebox{3.5ex}{\mbox{\bf IF~:}~~~~~}
a(-X \to -+X) & = \frac{1}{s_{AK}}\frac{y_{AK}^2}{y_{aj}} ~, \\
a(-X \to ++X) & = \frac{1}{s_{AK}}\frac{(1-y_{AK})^2}{y_{aj}} ~.
\end{align}

The helicity averages for $gX\to qqX$ antennae (gluon backwards evolving
to a quark) are
\begin{align}
\mbox{\bf II~:~~} a(g_A X_B \to q_a q_j X_b) & = \frac{1}{s_{AB}}\frac{1 +
  (1-y_{AB})^2}{2y_{aj} (1-y_{jb})}~, \\
\mbox{\bf IF~:~~} a(g_A X_K \to q_a q_j X_k) & = \frac{1}{s_{AK}}\frac{1
  + (1-y_{AK})^2}{2y_{aj} (y_{AK} + y_{aj})}~.
\end{align}

The individual helicity contributions are:
\begin{align}
a(+X \to ++X) & = \frac{1}{s_{AB}}\frac{1}{2y_{aj}(1-y_{jb})} ~, \\
a(+X \to --X) & = \frac{1}{s_{AB}}\frac{(1-y_{AB})^2}{2y_{aj}(1-y_{jb})} ~, \\
\raisebox{3.5ex}{\mbox{\bf II~:}~~~~~}
a(-X \to --X) & = \frac{1}{s_{AB}}\frac{1}{2y_{aj}(1-y_{jb})} ~, \\
a(-X \to ++X) & = \frac{1}{s_{AB}}\frac{(1-y_{AB})^2}{2y_{aj}(1-y_{jb})} ~.
\intertext{~}
a(+X \to ++X) & = \frac{1}{s_{AK}}\frac{1}{2y_{aj}(y_{AK}+y_{aj})} ~, \\
a(+X \to --X) & = \frac{1}{s_{AK}}\frac{(1-y_{AK})^2}{2y_{aj}(y_{AK}+y_{aj})} ~, \\
\raisebox{3.5ex}{\mbox{\bf IF~:}~~~~~}
a(-X \to --X) & = \frac{1}{s_{AK}}\frac{1}{2y_{aj}(y_{AK}+y_{aj})} ~, \\
a(-X \to ++X) & = \frac{1}{s_{AK}}\frac{(1-y_{AK})^2}{2y_{aj}(y_{AK}+y_{aj})} ~.
\end{align}

\subsection{Gluon Emission of Initial-State Gluons}
\label{app:GemiOffISG}

As discussed in \appsRef{app:QGemit} and \ref{app:GGemit},  helicity configurations exist for which a final-state gluon inherits the helicity of an initial-state gluon. Thus, the helicity of a pre-branching initial-state gluon, $A$ or $B$, can be different from that of the corresponding post-branching initial-state gluon, $a$ or $b$, without violating helicity conservation. 

For completeness, we give the DGLAP limits of antenna functions for gluon emission off initial-state gluons. 
The limits are independent of the other parent in the
parent antenna. For intial-initial antennae the DGLAP 
limit corresponds to
\begin{align}
  y_{jb}=\frac{Q^2}{s_{ab}}\to 0~,~~z=y_{AB}=\frac{s_{AB}}{s_{ab}}~,~~\text{and}
  ~~y_{aj}\to1-z~.
\end{align}
This gives the following limits of the helicity-dependent antenna functions in
\eqsRef{eq:GGemitIIhelFirst} to \eqref{eq:GGemitIIhelLast} (or 
\eqsRef{eq:QGemitIIhelFirst} to \eqref{eq:QGemitIIhelLast}) for a parent with
$+$ helicity,
\begin{alignat*}{3}
a(X+\to X++) & ~\to~ \frac{1}{Q^2}\,\frac1z\,\frac{1}{z(1-z)} 
  & ~=~ & \frac{1}{Q^2}\,P_{g\to gg}^\mrm{IS}(+\to ++)~, \\     
a(X+\to X-+) & ~\to~ \frac{1}{Q^2}\,\frac1z\,\frac{z^3}{1-z} 
  & ~=~ & \frac{1}{Q^2}\,P_{g\to gg}^\mrm{IS}(+\to -+)~, \\
a(X+\to X--) & ~\to~ \frac{1}{Q^2}\,\frac1z\,\frac{(1-z)^3}{z} 
  & ~=~ & \frac{1}{Q^2}\,P_{g\to gg}^\mrm{IS}(+\to +-)~.
\end{alignat*}
The same limits are obtained for initial-final antennae with
\begin{align}
  y_{aj}=\frac{Q^2}{m_{AK}^2+s_{jk}}\to 0~,~~z=y_{AK}=
  \frac{m_{AK}^2}{m_{AK}^2+s_{jk}}~,~~y_{jk}\to1-z~,~~\text{and}
  ~~y_{ak}\to1~.
\end{align}

\section{Details of VINCIA Implementation \label{app:implementation}}
For completeness, we also report on the following changes in \vc with respect to \cite{Fischer:2016vfv}:
\begin{newItem}
\item The so-called ``Ariadne factor''~\cite{Lonnblad:1992tz} for gluon splitting antennae has been removed completely, as it has only been applied to 4-jet events in hadronic $Z$ decays and its influence cancels once ME corrections are used in the evolution.
\item The CMW-rescaling of $\alpha_s$~\cite{Catani:1990rr} is no longer applied to the soft-eikonal terms of the antenna functions, but rather as a global rescaling of $\Lambda_\mrm{QCD}$, independent 
of the type of branching.
\item By default the power shower approach~\cite{Miu:1998ju,Plehn:2005cq} is used for hard process without QCD partons in the final state. This obviates the need for a separate event sample containing jets associated with scales
larger than the factorisation scale, which has been introduced in~\cite{Fischer:2016vfv}. For 
QCD-type processes the shower starts the evolution at the factorisation scale.
\item The so-called ``smooth ordering''~\cite{Giele:2011cb}, which allows the shower to populate phase-space regions
beyond the reach of traditional ordered showers, is no longer used. Consequently, the MECs formalism
so far used in \vc is no longer applicable and the MECs method for ordered showers 
of~\cite{Fischer:2017yja} is applied. See \secRef{sec:mecs} for a brief review of the formalism.
\item The CKKW-L merging implementation in \py8~\cite{Lonnblad:2011xx} is now also available in \vc, making use of the parameters in \py8. This allows to supplement the MECs method for ordered 
showers with non-shower-like events, as discussed in~\cite{Fischer:2017yja}. Note however, that it 
is not possible to combine the merging procedure with the helicity-dependent shower.
\item Automated uncertainty variations are now user specifiable in the same manner as in \py~8~\cite{Mrenna:2016sih}, with the following keywords controlling the type and size of variations in \vc, for final-final (FF), initial-final (IF), and initial-initial (II) antennae respectively:
\begin{itemize}
    \item Renormalisation-scale variations (applied to all antenna functions): \begin{center}
    \ttt{ff:muRfac}~;~ \ttt{if:muRfac}~;~ \ttt{ii:muRfac}~.
    \end{center}
    \item Nonsingular-term variations (applied to all antenna functions): 
    \begin{center}
    \ttt{ff:cNS}~;~ \ttt{if:cNS}~;~ \ttt{ii:cNS}~. 
    \end{center}
    \item Optionally,  antenna-specific variations 
    can be specified, which then supersede the antenna-independent variations. The full set of possible keywords is listed in \vc's HTML User Reference. 
    \item As an example, the following value (default in \vc 2.200) for the \ttt{Vincia:UncertaintyBandsList} string vector
    defines a set of four alternative weight sets, with labels 
    ``alphaShi'', ``alphaSlo'', ``hardHi'', and ``hardLo'', respectively: 
\begin{verbatim}
 Vincia:UncertaintyBandsList = {
    alphaShi ff:muRfac=0.5 if:muRfac=0.5 ii:muRfac=0.5, 
    alphaSlo ff:muRfac=2.0 if:muRfac=2.0 ii:muRfac=2.0, 
    hardHi ff:cNS=2.0 if:cNS=2.0 ii:cNS=2.0, 
    hardLo ff:cNS=-2.0 if:cNS=-2.0 ii:cNS=-2.0 }
\end{verbatim}
\end{itemize}
\end{newItem}

\bibliography{main}
\bibliographystyle{JHEP}

\end{document}